# The Shifts Hypothesis - an alternative view of global climate change

Belolipetsky P.V. [1,2]

[1] *Institute of computational modelling, SB RAS, Krasnoyarsk, Russia*

[2] *Institute of biophysics, SB RAS, Krasnoyarsk, Russia*

*email: pbel@icm.krasn.ru*

## Abstract

In this study we used HadCRUT4 monthly mean temperature anomalies for 1950-2013 years in order to investigate properties of recent warming. Our aim was to separate changes produced by short-term ENSO variations and to look on temporal and spatial dynamics of residual temperature anomalies. For this we subtract linear influence of ENSO index from each HadCRUT4 grid box. We found that residual global temperature dynamics looks like staircase function: linear trends for three quasi-stable periods 1950-1987, 1988-1997 and 1998-2013 are near zero and near all warming occurred during two shifts of 1987/1988 and 1997/1998 years. The maps of linear trends during quasi-stable periods suggest that there are no any global changes and observed variations produced by local changes. In contrast majority of regions quickly warmed near 1987/1988 and 1997/1998 shifts.

All these allows us to formulate a new hypothesis about recent warming - the Shifts Hypothesis. It explains the structure of recent warming as follows: during shifts of 1987/1988 and 1997/1998 the mean value of global temperature quickly rose, over which natural variability associated with ENSO, PDO and many other local factors occurs.

The Shifts Hypothesis have advantages over other hypotheses in terms of simplicity. Its simplicity is that it uses only two factors to obtain an explanation of general features of global temperature anomalies dynamics in 1950-2013. This, for example, allows to build very simple linear regression models allowing with only two factors (ENSO and climate regimes) adequately reproduce observed dynamics and account for 78% in monthly mean and 87% in yearly mean global temperature anomalies.

According to IPCC forcing-response paradigm warming process (WP) cleared from all factors should be continuous warming (may be near linear on the considered time interval). This idea is inconsistent to the obtained staircase function. This staircase function may be explained by current IPCC forcing-response paradigm only if it happened so that the combination of compensating factors is near the same as anthropogenic greenhouse gases (GHGs) forcing for each quasi-stable period - 1950-1987, 1988-1997 and 1998-2013. This seems very unlikely. An alternative and more simple explanation is the existence of some regulation mechanism (e.g. global thermostat) not presented in IPCC climate models. This mechanism should maintain global temperature near stable in 1950-1987, 1988-1997 and 1998-2013 periods nevertheless all the time growing forcing due to anthropogenic GHGs. There is the same situation with 1987/1988 and 1997/1998 shifts. What explanations can be provided? According to IPCC paradigm these shifts may be only the result of coincidence, which seems very unlikely. And the simplest explanation is that sometimes some parameter of global thermostat can change that lead to a new normal global mean temperature (or setpoint of thermostat).

## Adjustment for ENSO

Let's consider recent climate change as a combination of different processes. Suppose that temperature anomalies T(lat, lon, t) are linear superposition of two factors - quasi-periodic natural influences NV(lat, lon, t) and some warming process WP(lat, lon, t). So

$$T(lat, lon, t) = NV(lat, lon, t) + WP(lat, lon, t)$$

This warming process may be natural or anthropogenic - we don't know. We even don't know what variations of T are due to WP and what are due to NV. But in order to understand the mechanisms of WP it is useful to estimate the temporal evolution (e.g. WP(t)) and spatial pattern (e.g. WP(lat, lon, t)). From 1950 there is near global coverage of surface temperature measurements, that is provided by HadCRUT4 dataset. Thus we can suppose that we know quiet enough about temperature anomalies T(lat, lon, t). So in order to obtain WP(lat, lon, t) we should subtract natural variations from observed temperature anomalies:

$$WP(lat, lon, t) = T(lat, lon, t) - NV(lat, lon, t).$$

Unfortunately we don't know exactly, how NV(lat, lon, t) looks like. It may be for example some combination of ENSO, PDO, AMO, QBO, volcanic and solar forcing. So it is very complicated task to fully reconstruct NV(lat, lon, t).

By lack there is large reconstructable part in NV(lat, lon, t) - simultaneous (not lagged) linear part of ENSO. Let's perform empirical orthogonal function (EOF) analysis of HadSST3 dataset by instruments available at Climate Explorer site (climexp.knmi.nl). What EOF analysis do? In general it is seeking by some algorithm spatial functions EOF(lat, lon) and time-series so-called principal components PC(t) in order to minimize residuals RES in a linear decomposition:

$$HadSST3(lat,lon,t) = EOF1(lat,lon)*PC1(t) + EOF2(lat,lon)*PC2(t) + ... + EOF2(lat,lon)*PC2(t) + RES(lat,lon,t)$$

Let's look on first empirical orthogonal of HadSST3:

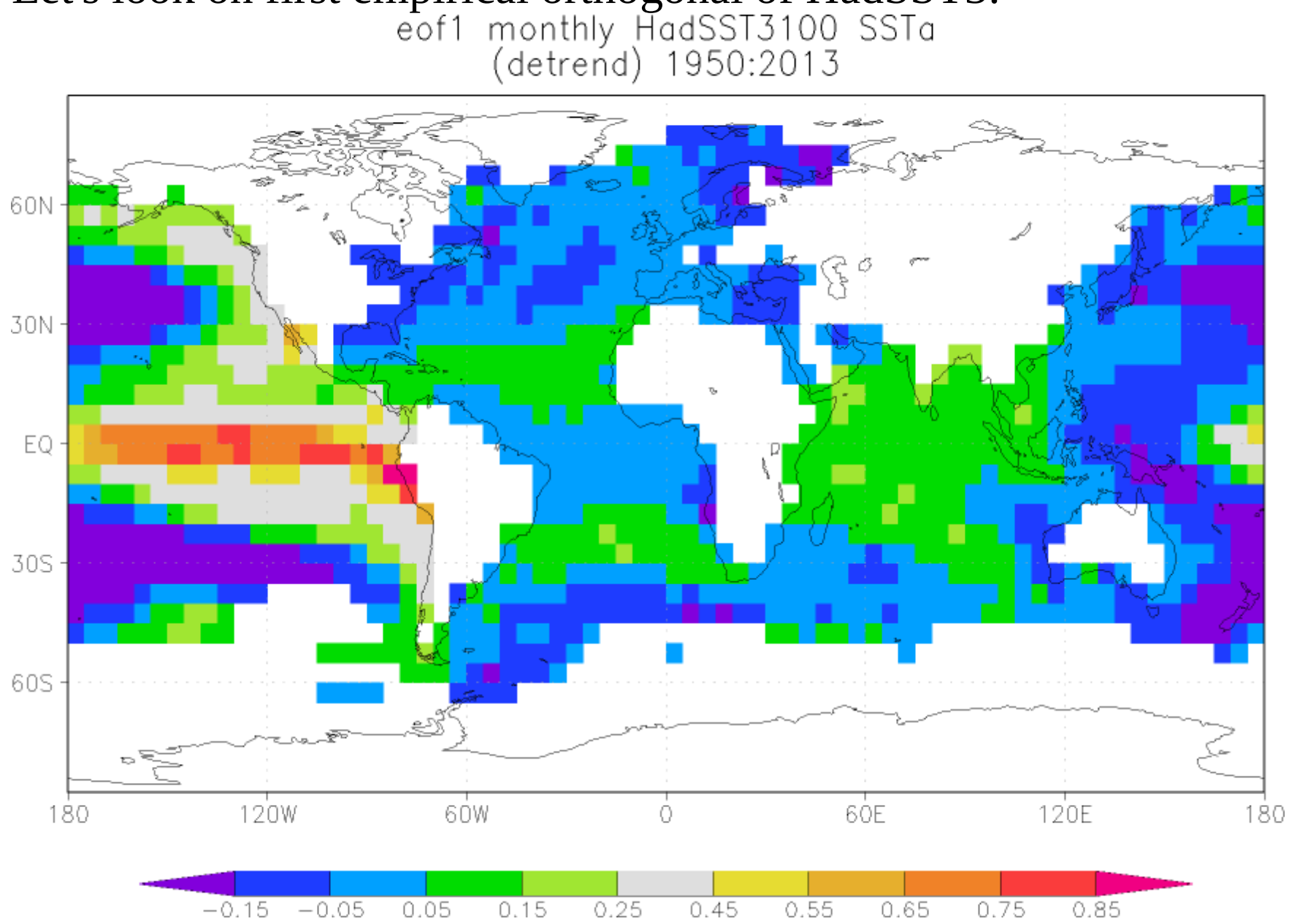

Fig. 1. Map of first empirical orthogonal function of HadSST3 for 1950-2013 years.

It is showing well described effects of ENSO - during El Nino central east Pacific is significantly warmer then normal, surrounding areas of Pacific are slightly cooler then normal and Indian ocean is slightly warmer then normal. And a first principal component PC1(t) is near the same as various ENSO indexes:

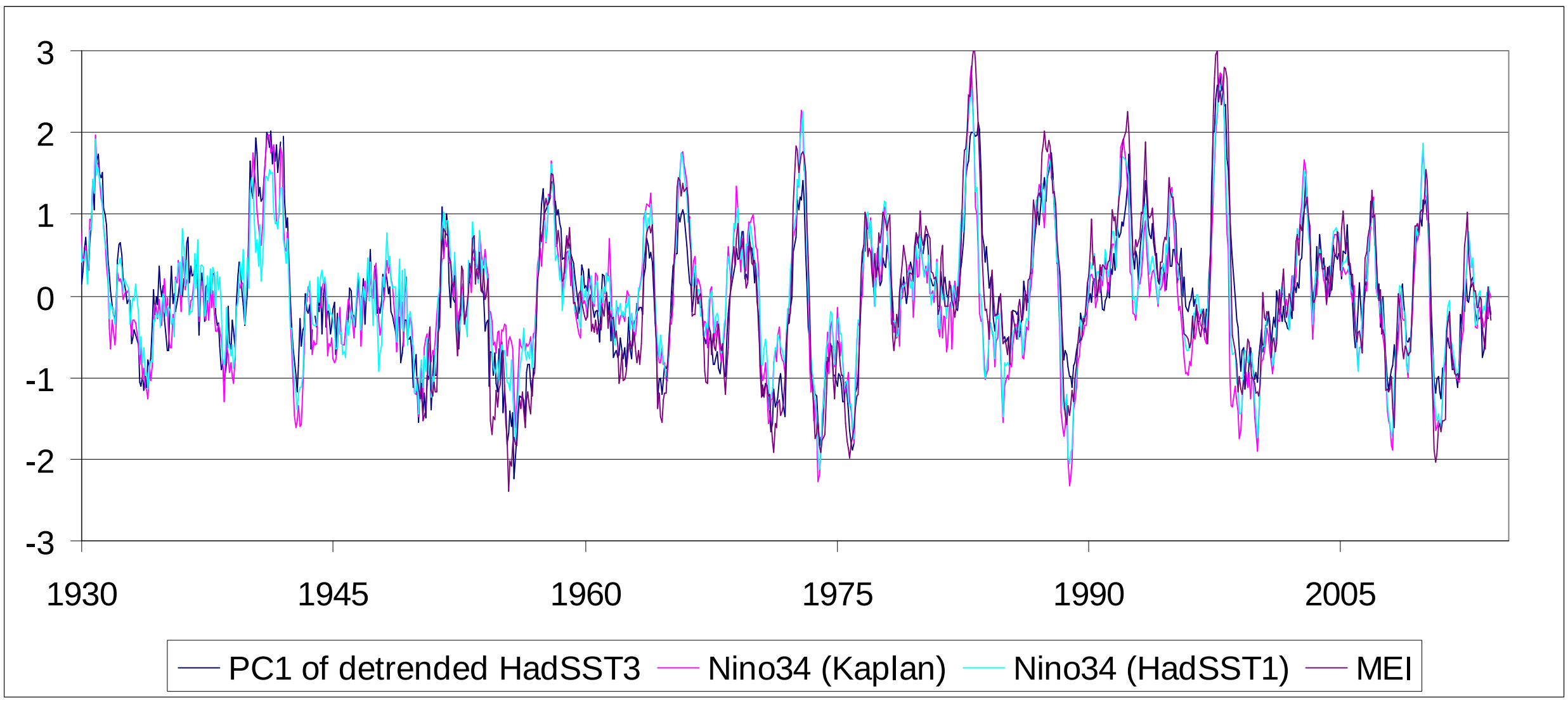

Fig. 2. Comparison of different ENSO indexes.

Spatial pattern of the first EOF and temporal evolution of the first PC is showing that there is quasi-periodical short-term process producing large changes to temperature anomalies field T(lat, lon, t). Obviously this process is ENSO and adjustment for it is needed in order to understand spatial-temporal evolution of warming process WP(lat, lon, t). For this adjustment we will use first PC of detrended HadSST3 instead of other popular indexes, because indexes like Nino34 and SOI are produced by measurements at local positions and more susceptible to local factors.

It may be near impossible to fully remove the influence of ENSO from temperature anomalies dataset. So our aim is to adjust for most big variations obviously produced by ENSO and to minimize the risk of introducing biases. We argue that the most simplest and unbiased method is to subtract linear influence of our ENSO index from each grid cell of HadCRUT4. By this procedure we don't mix regions with high and low ENSO influence (in contrast to consideration of global averages). Also we don't diminish the amplitude of warming because the linear trend of ENSO index is near zero. We performed this procedure by means of Climate explorer site. See the step-by-step description in supplementary materials. Commonly it takes less then 10 minutes to reproduce the results.

In order to illustrate how the adjustment works lets consider the temperature anomalies of cell with coordinates from 0N to 5N and from 125W to 120W. It is a cell in a Nino34 region and monthly temperature anomalies there look so:

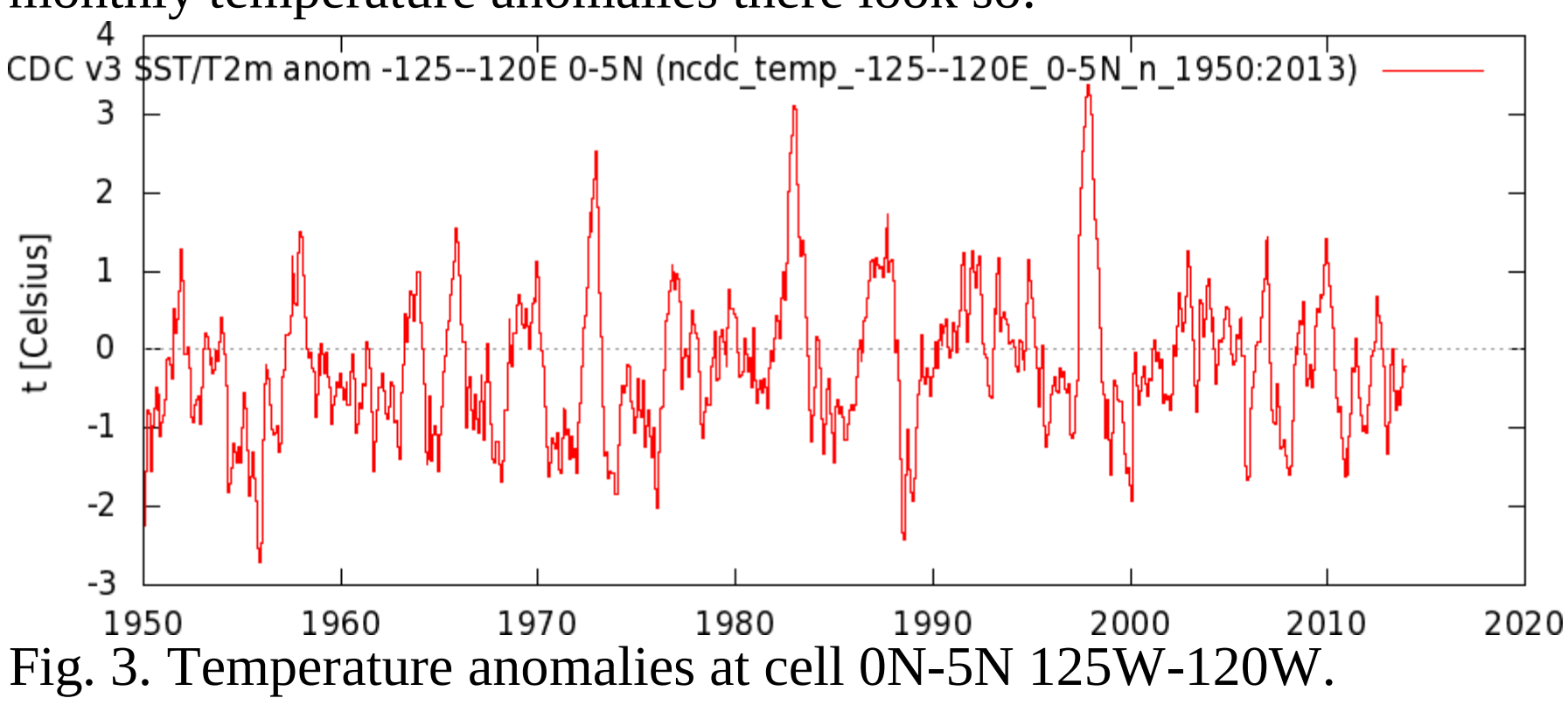

Fig. 3. Temperature anomalies at cell 0N-5N 125W-120W.

From the other hand we have time series of our ENSO index (first principle component of HadSST3):

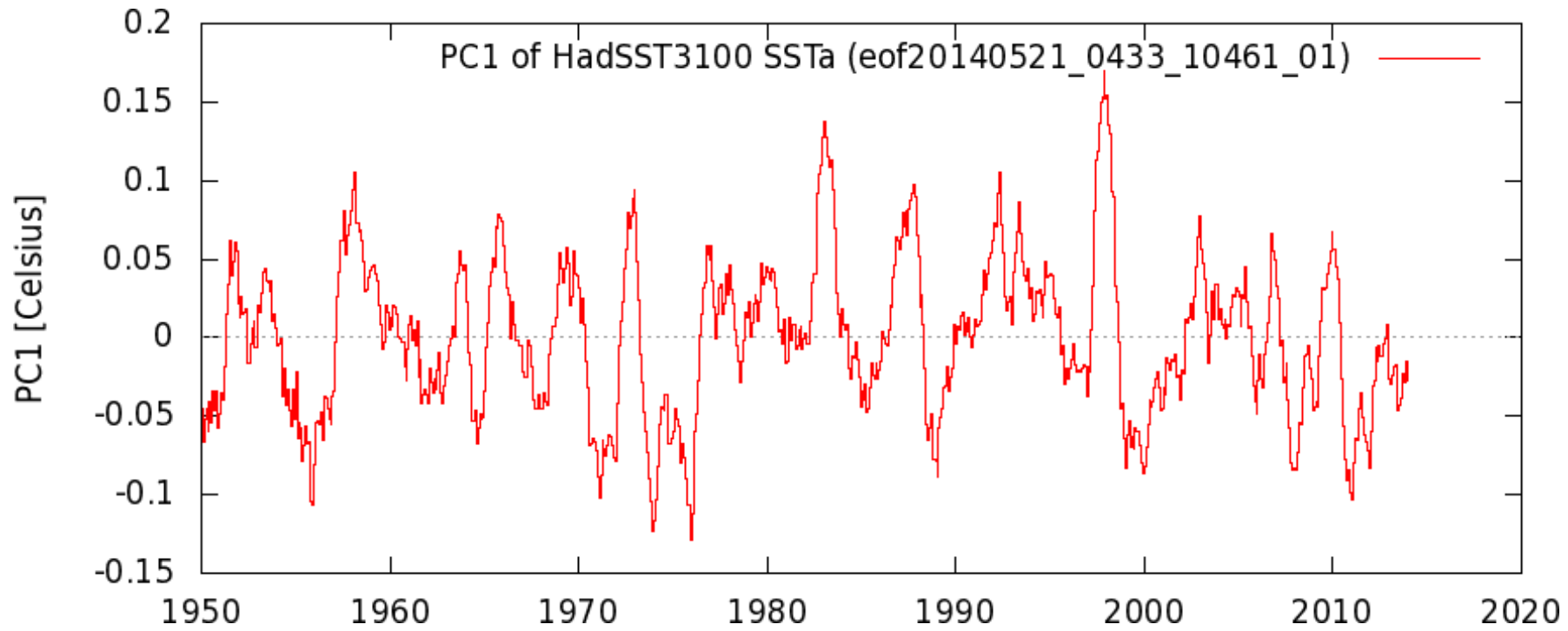

Fig. 4. Our ENSO index.

If we plot this time series on one graph they will be looking so:

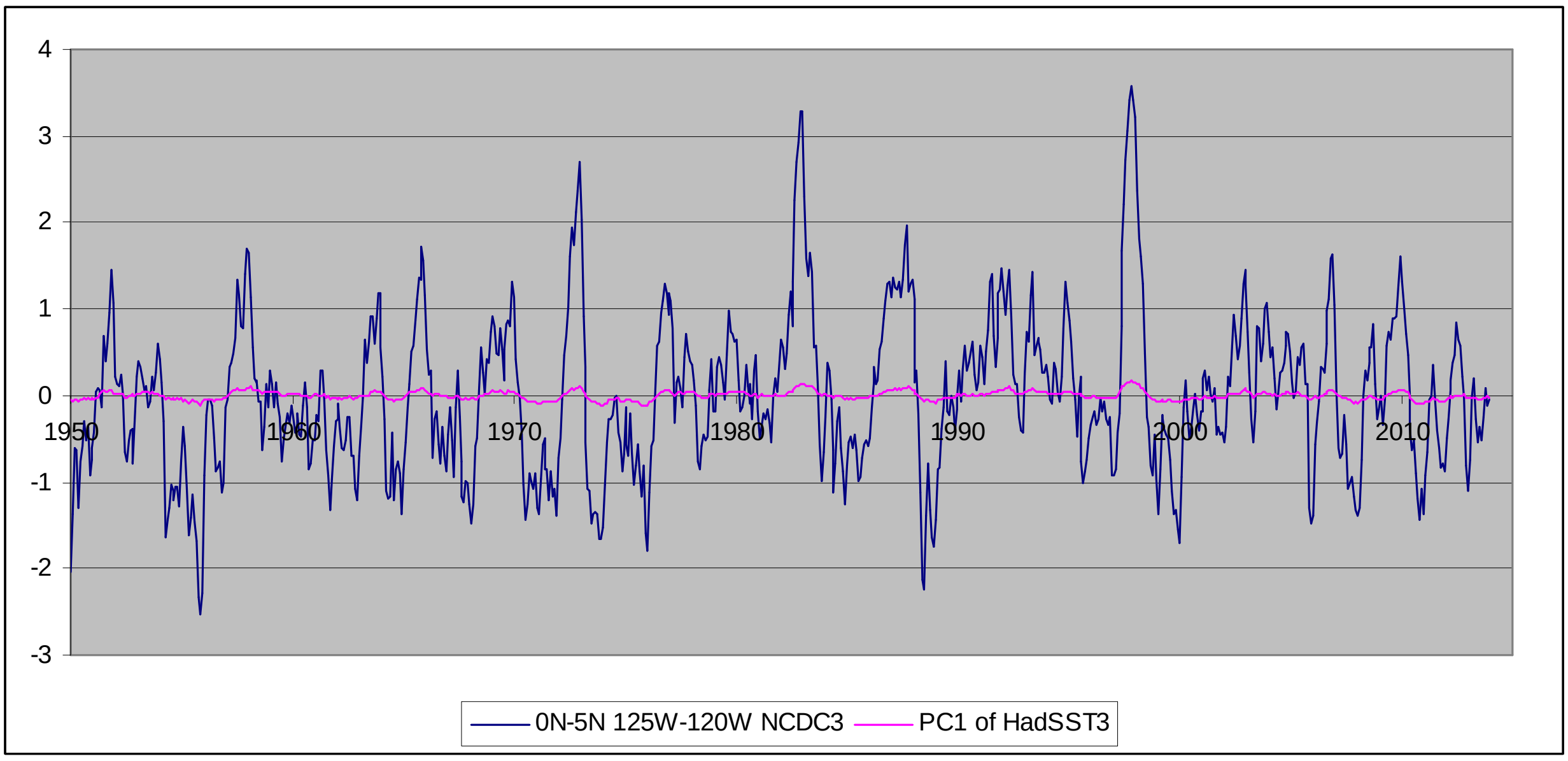

Fig. 5. Comparison of temperature anomalies at cell 0N-5N 125W-120W and our index.

Now we can perform linear regression of 0N-5N 125W-120W NCDC3 (let's call it T(t)) on PC1 of HadSST3 (let's call it ENSO(t)):

$$T(t) = a*ENSO(t) + b + Residual(t).$$

For example, we can do it by means of Excel and obtain the following values a=15.99, $b=4*10^{-7}$. So a*ENSO(t) we consider as linear influence of ENSO. A comparison of T(t) and a*ENSO(t) can be seen on the next figure:

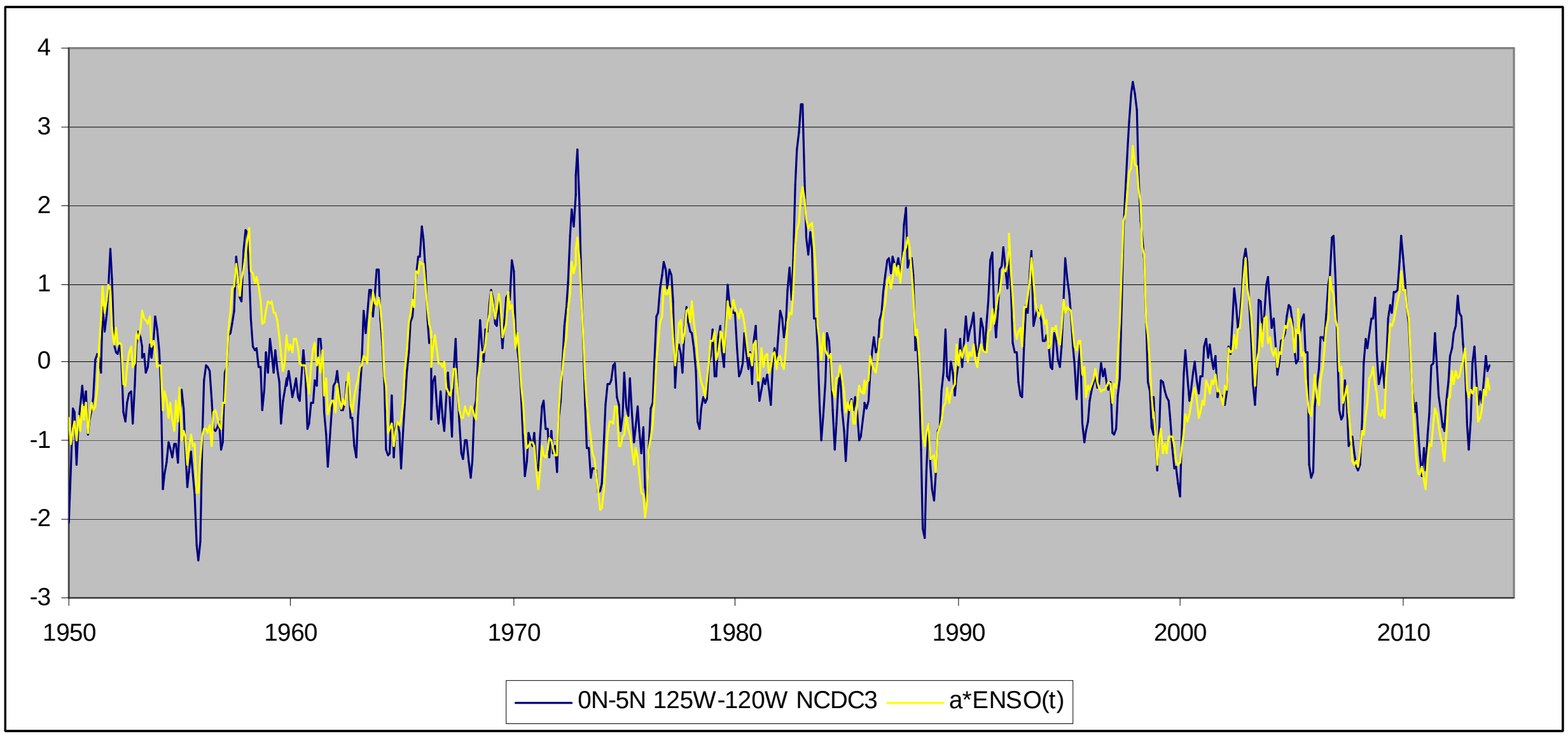

Fig. 6. Comparison of temperature anomalies and linear ENSO effect on them.

Now I think it is clearly seen what part of variations at this locations are due to ENSO. Also I think some no-linear effects of ENSO are also visible: during strong El Ninos of 1972, 1983 and 1998 temperatures here rose more then linearly; the same is observed during some La Ninas.

In order to adjust temperature anomalies at this location for linear ENSO effects we should subtract a*ENSO(t) from T(t). As a result we obtain adjusted T(t) and it can be compared with initial:

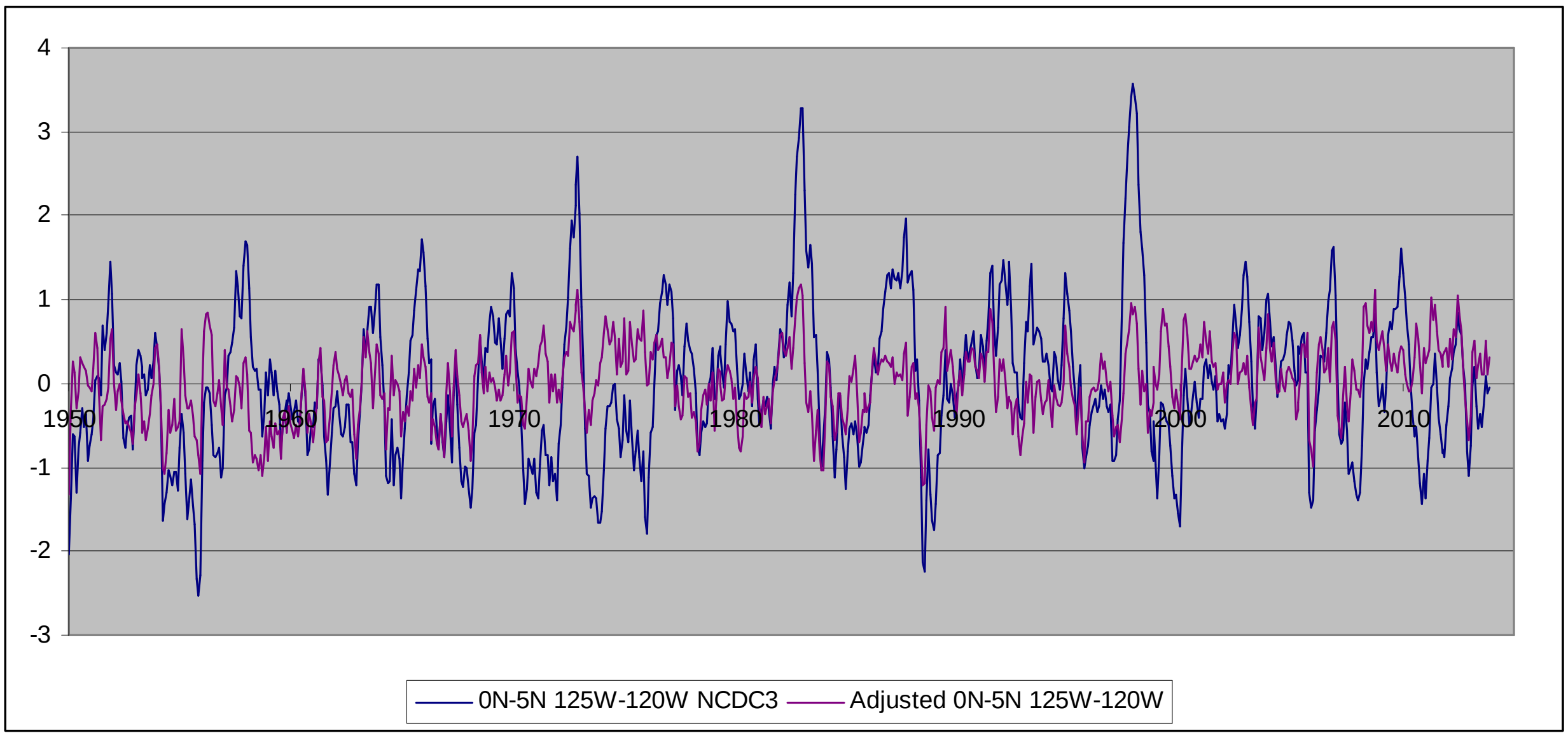

Fig. 7. Comparison of initial and adjusted time-series.

There still some residual of ENSO influence, but the largest part of it was removed. See excel file with calculations for this cell in supplementary materials.

Climate Explorer allows to perform this procedure for each grid cell of HadCRUT4 by several actions (steps 7-11 at Description of ENSO adjustment). As a result we obtain a new field adjusted_HadCRUT4(latitude, longitude, time) which we can visualize and investigate by means of Climate Explorer. Also we can download this field or time series for chosen region and analyze it in another software. For example, for me it is convenient to draw plots in Excel.

Let's now look on changes which are produced by this adjustment to different regions. Most changes produced by this simple adjustment are located in tropical east Pacific:

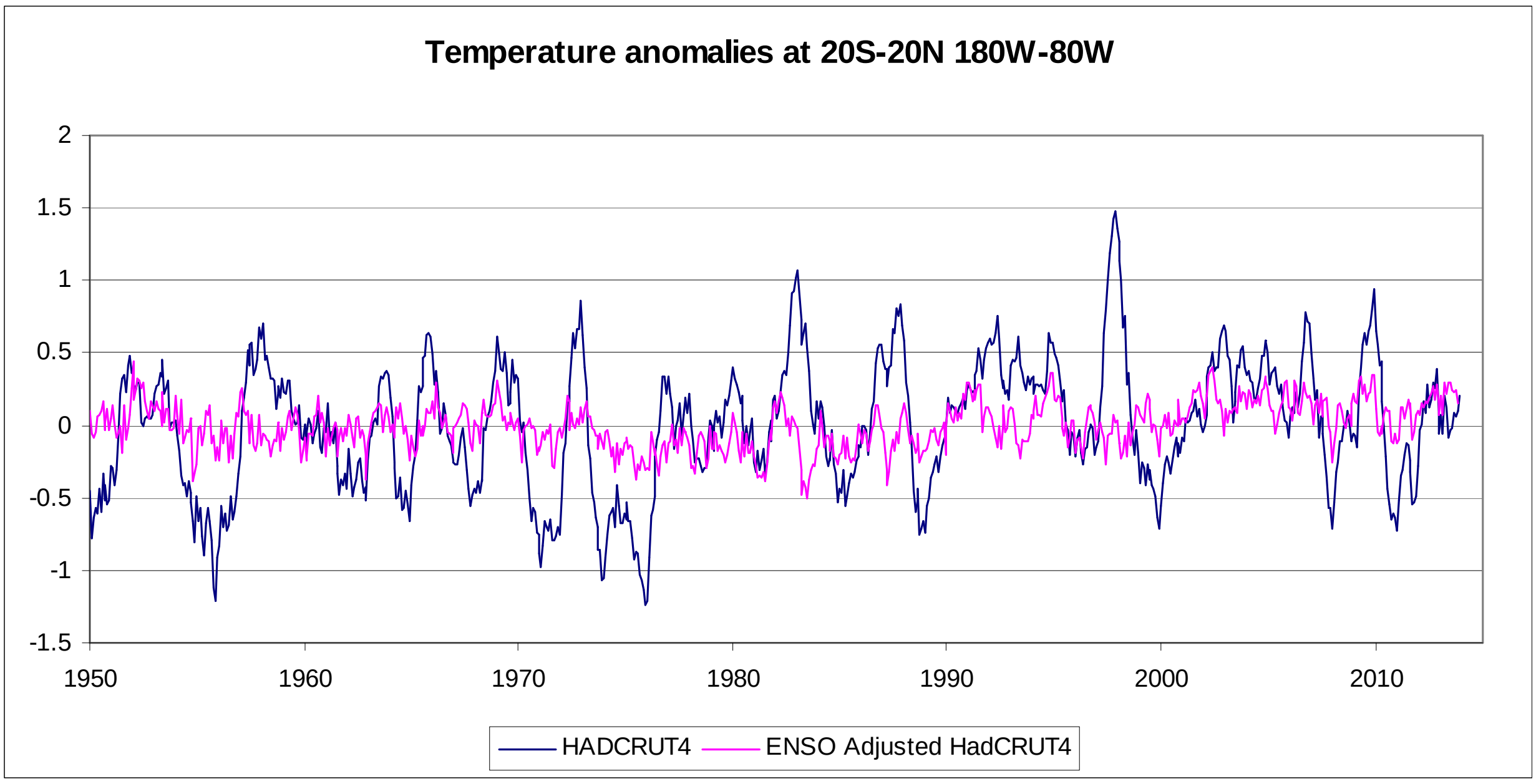

Fig. 8. Comparison between initial HadCRUT4 temperature anomalies and after adjustment at tropical east Pacific.

Adjustment in Indian ocean are not so good, changes are smaller then needed:

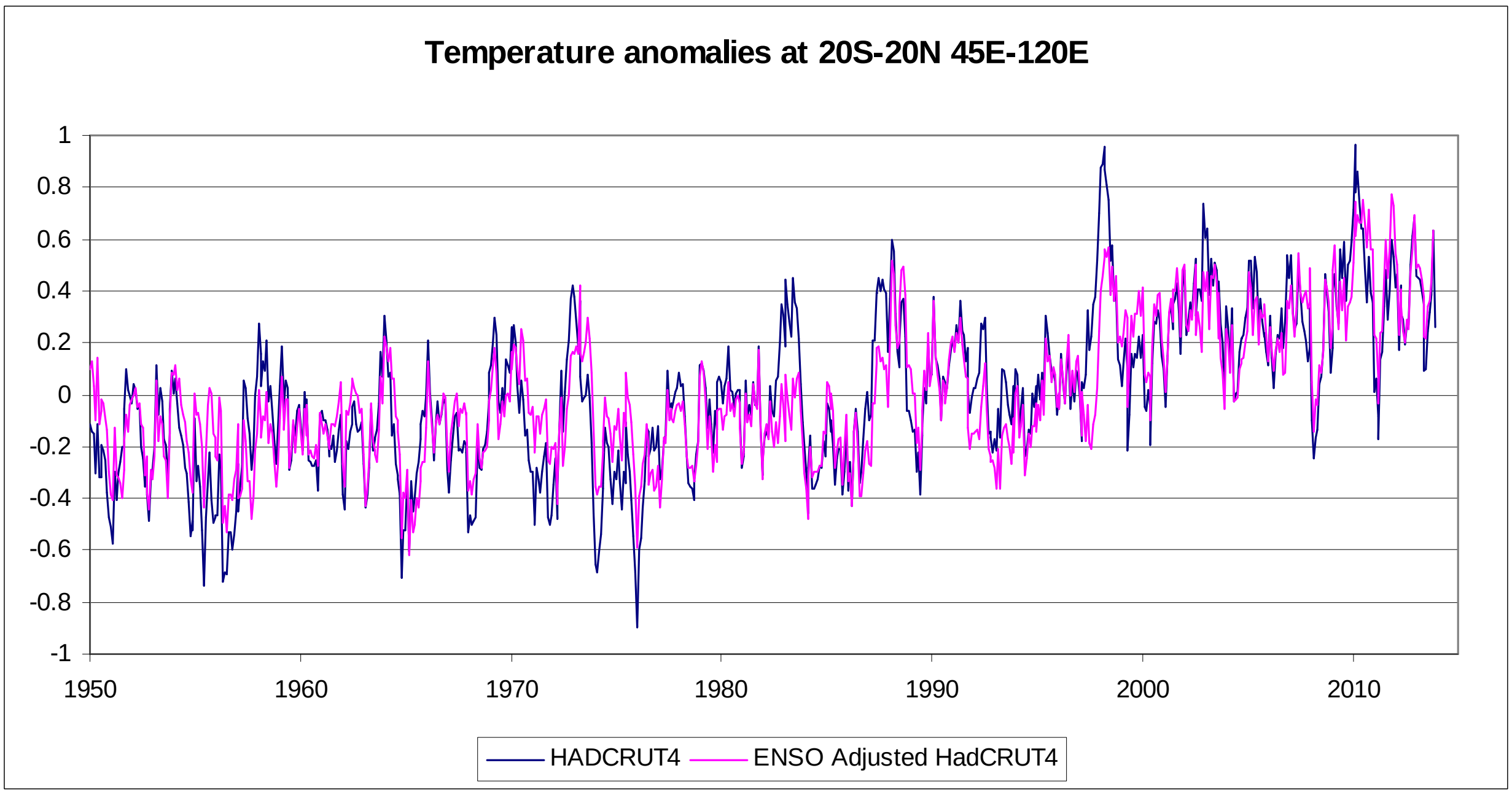

Fig. 9. Comparison between initial HadCRUT4 temperature anomalies and after adjustment at tropical Indian ocean.

In other regions changes are very small. North extra tropical latitudes looks so:

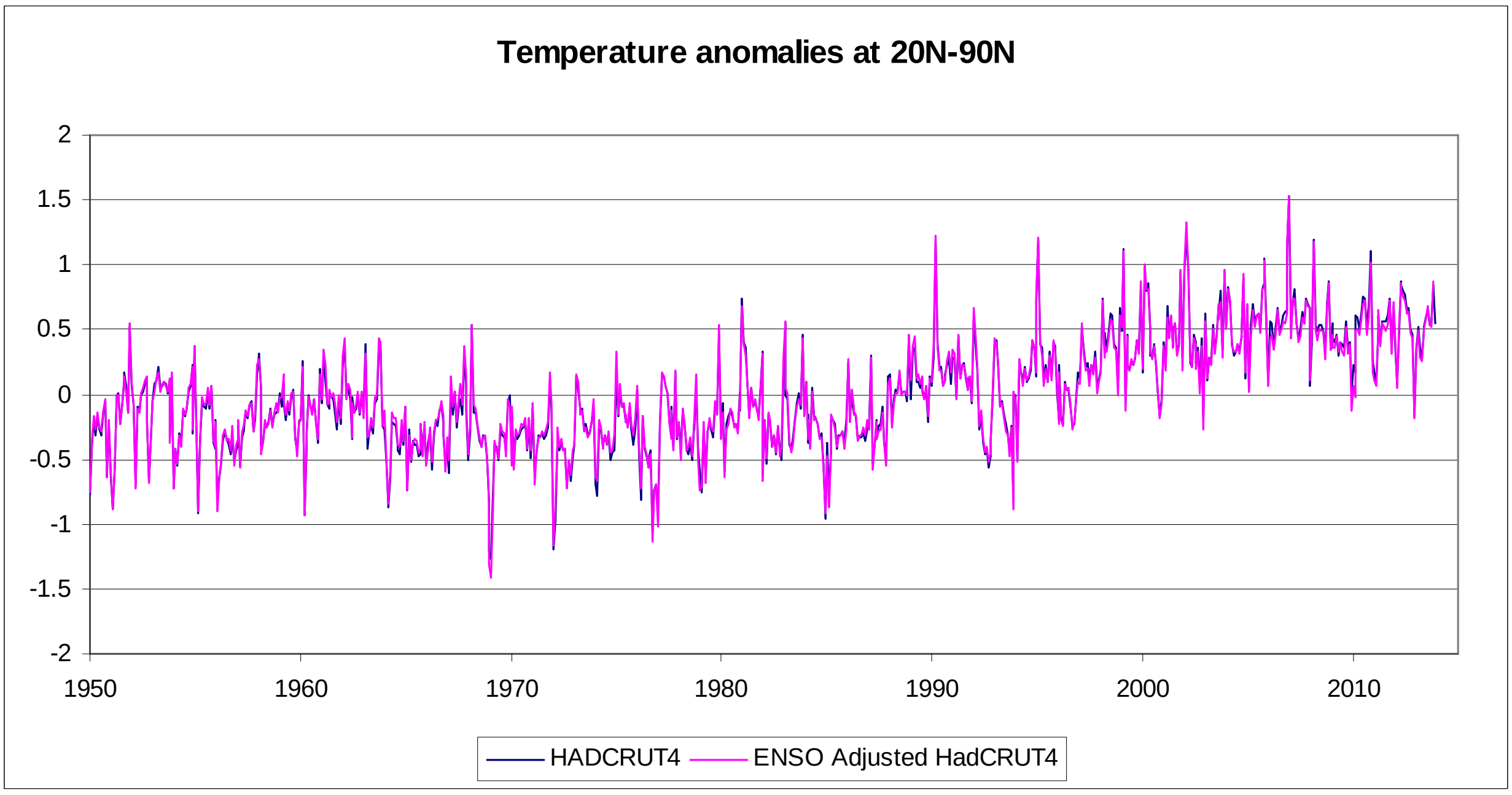

Fig. 10. Comparison between initial HadCRUT4 temperature anomalies and after adjustment at north extra tropical latitudes.

South extratropical latitudes looks so:

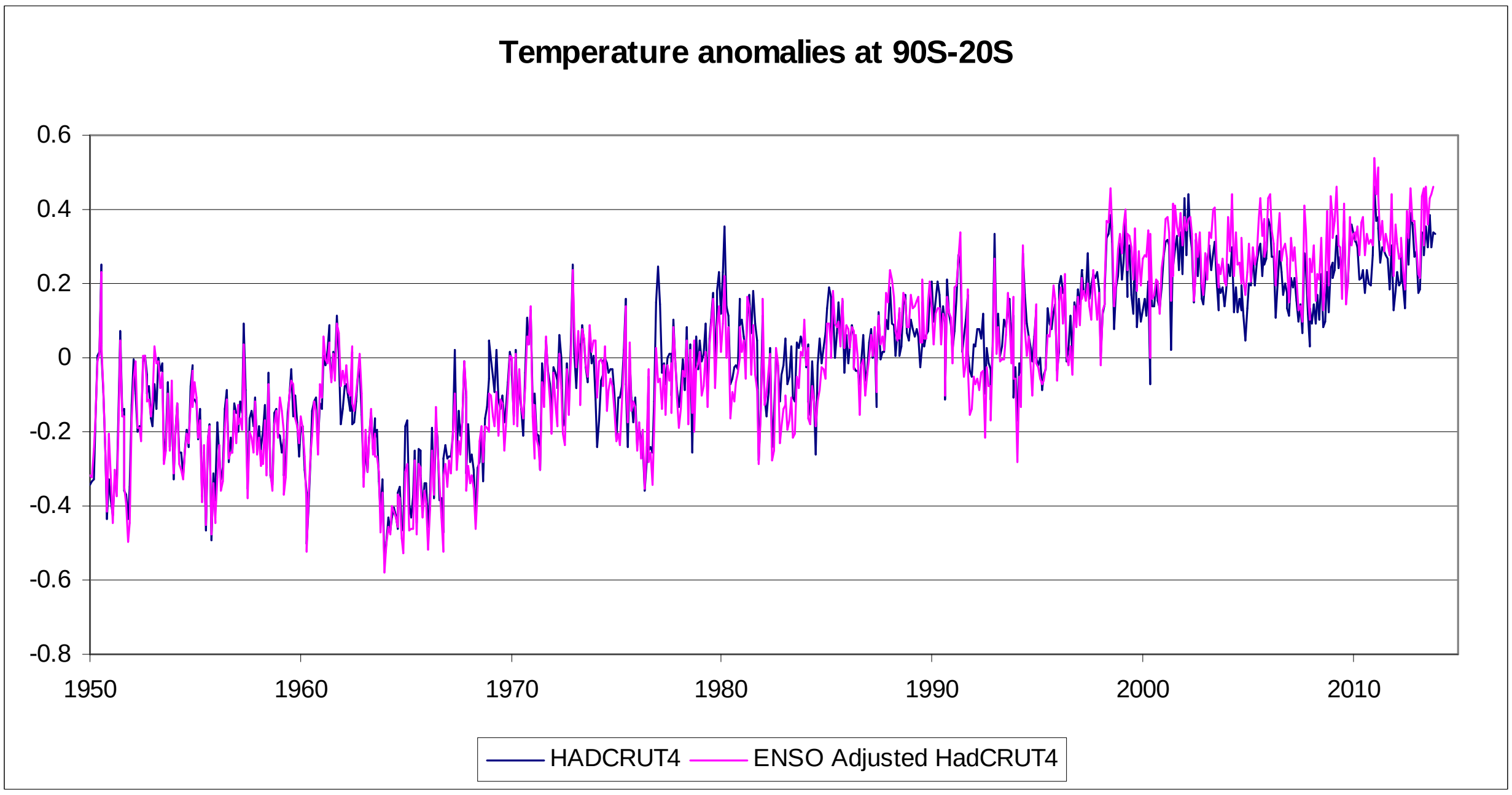

Fig. 11. Comparison between initial HadCRUT4 temperature anomalies and after adjustment at south extra tropical latitudes.

Another way to get view of produced by adjustment changes is to look on a correlation map between initial HadCRUT4 and after adjustment. In most places correlation between initial and adjusted temperature anomalies is higher then 0.98. And it is clearly seen that most changes are in tropical east Pacific.

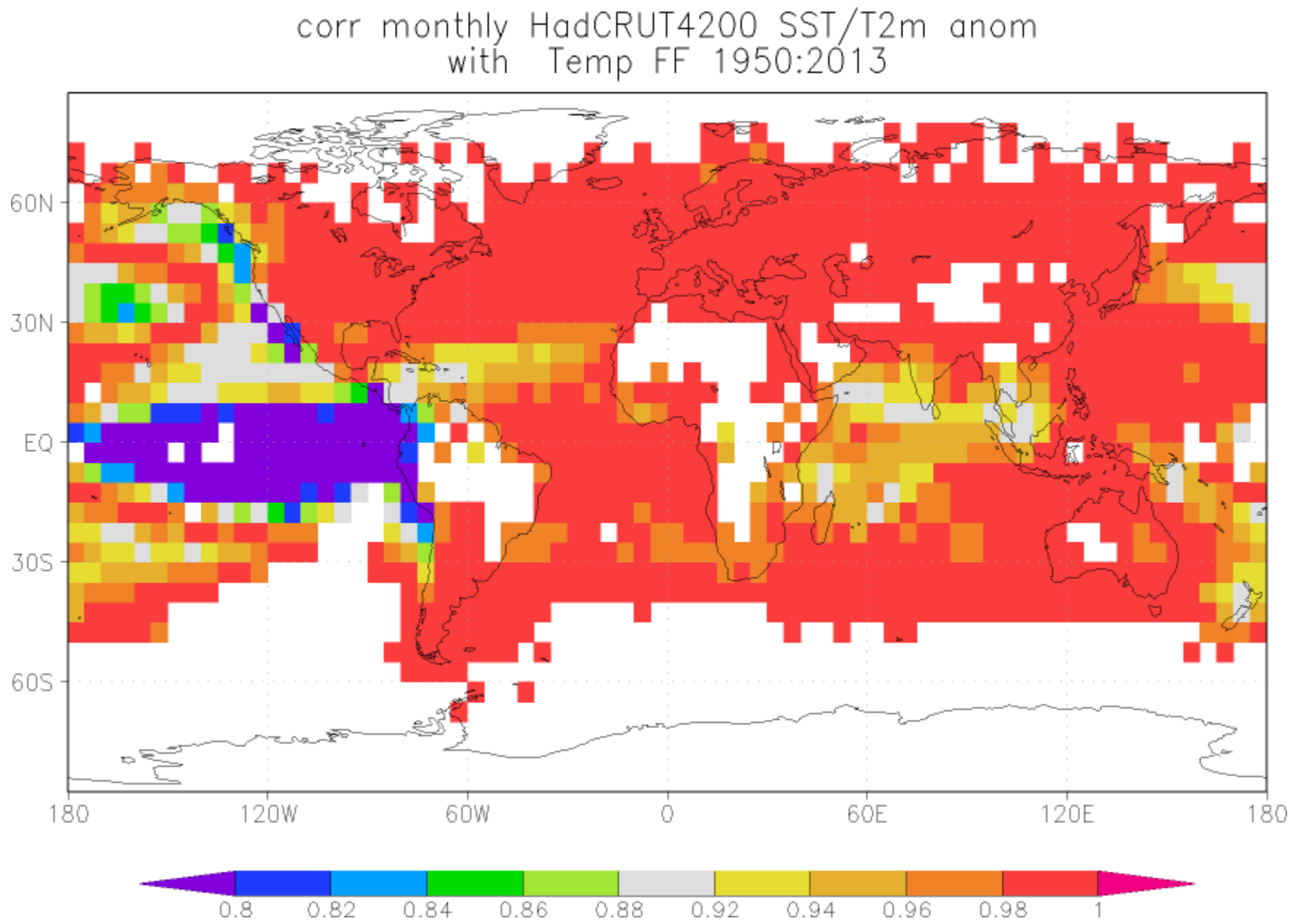

Fig. 12. Map of correlations between initial HadCRUT4 temperature anomalies and after adjustment.

So what we are doing and what is the meaning? We have obtained a first approximation of WP by performing:

WP1(lat,lon,t) = T(lat, lon, t) - ENSO_Lin_Eff(lat, lon, t),

there ENSO_Lin_Eff - are linear not lagged effects of ENSO. Main effect of our adjustment is that there are no more huge warming of Central East Pacific during El Nino and huge cooling during La Nino. Also there are some changes in other regions of world. We got only a first approximation of WP because besides of ENSO_Lin_Eff there are other components in natural influences NV - PDO, AMO and etc, and also nonlinear ENSO effects, lagged ENSO effects (e.g consequences of El Nino and La Nino). We are not talking now about all this components. What we have done is that we simply subtract **really existent** linear not lagged ENSO effects. During this procedure it doesn't matter what is a function of ENSO (e.g. a heat pump to remove heated water from the Pacific and move it to the poles), it doesn't mater what are the mechanisms (e.g. Ekman transport and Kelvin waves and the effects of sloshing water in the Pacific basin and the influence of the Madden-Julian Oscillation and the like). **For us it is important that we are subtracting really existent short term natural variations and can see better how warming process looks.**

After performed procedure we can look at global monthly mean temperature anomalies adjusted for linear ENSO effects:

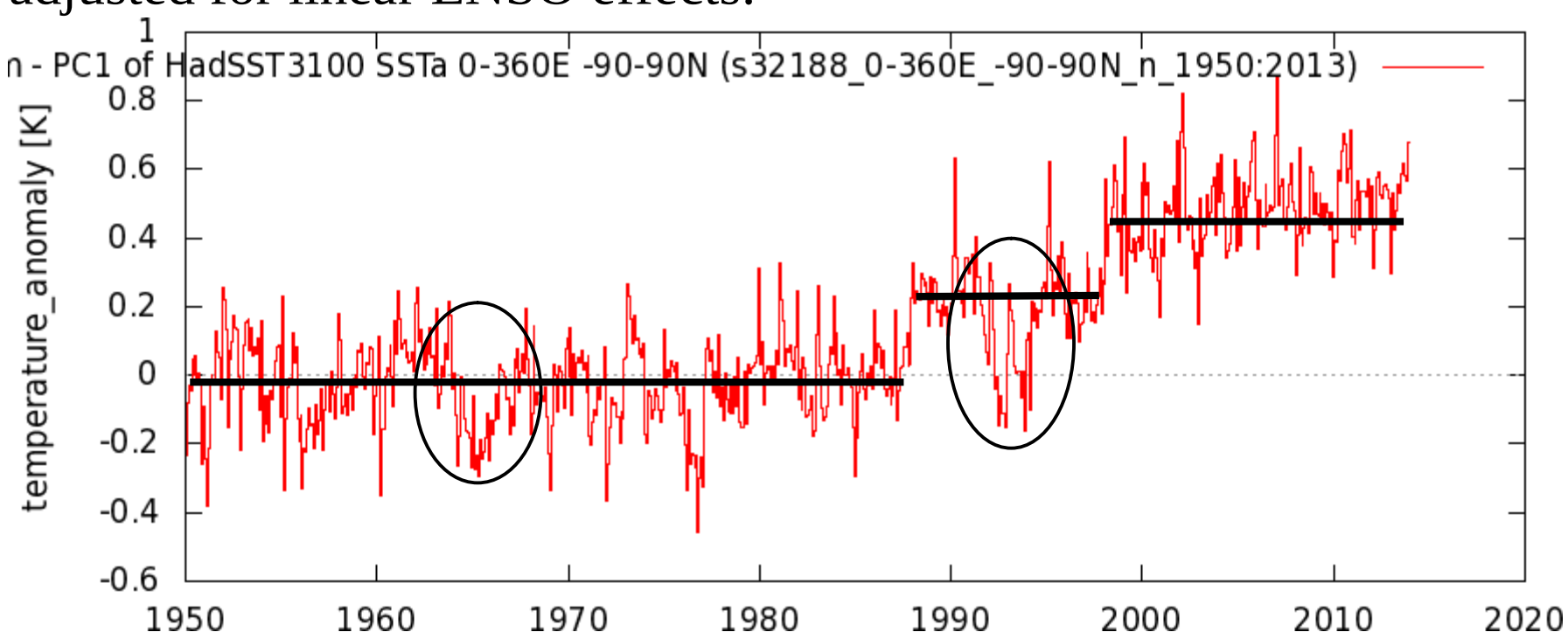

Fig. 13. Global monthly mean temperature anomalies of adjusted for ENSO HadCRUT4.

Linear trends during three quite stable periods illustrated by black lines are near zero. The main disturbances of black lines (highlighted by ellipses) are likely effects of Pinatubo and Agung volcanic eruptions. Especially strong perturbation observed after Pinatubo eruption at 1991. So the next step should be:

## Adjustment for volcanic eruptions

Let's look at global stratospheric aerosol optical depth due to volcanic eruptions from 1950 till now reconstructed by NASA\GISS:

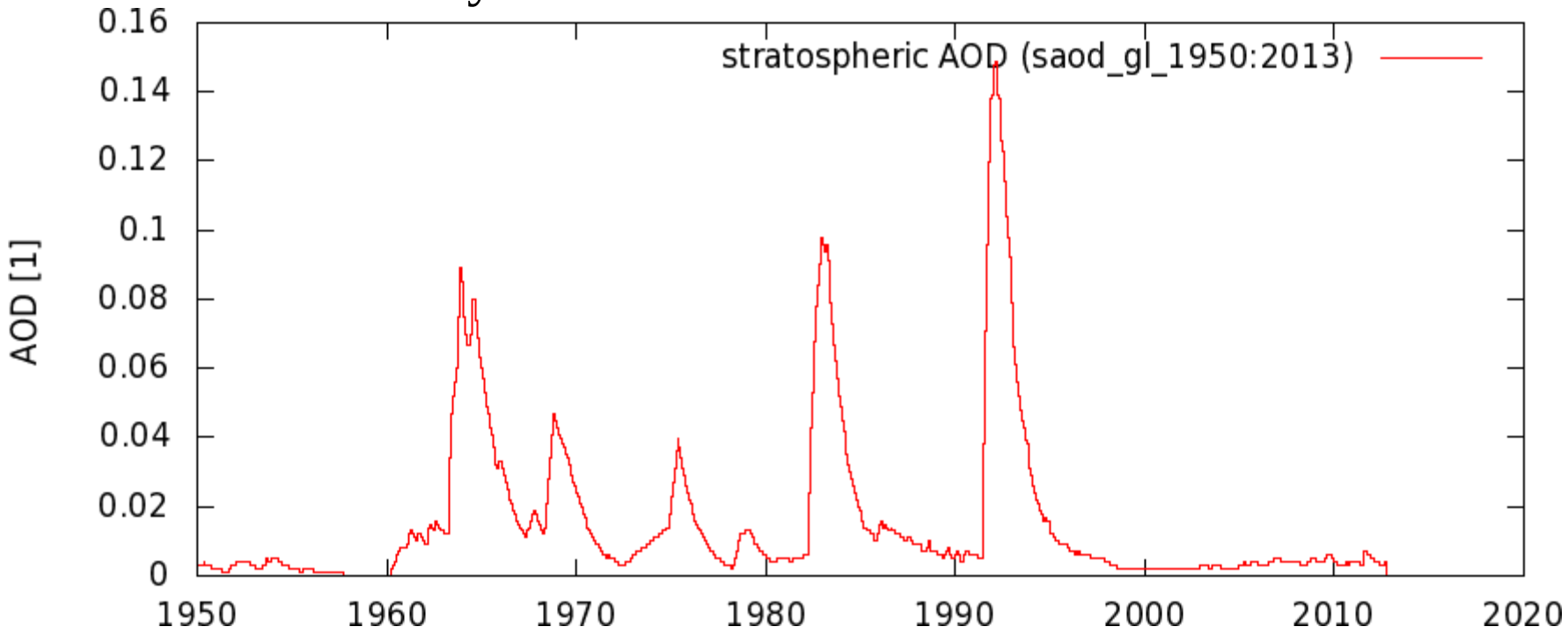

Fig. 14. Stratospheric aerosol optical depth from Climate Explorer reconstructed by NASA\GISS

There are three clearly distinguishable largest volcanic eruptions: Agung in 1963, El Chichon in 1982 and Pinatubo at 1991. Then again let's look at global monthly mean temperature anomalies with linear ENSO effects removed (Fig. 9). Effects of Agung and Pinatubo are clearly seen and practically there is no effect of El Chichon eruption. I was thinking about what is the difference between Agung and Pinatubo from one side and El Chichon from the other. It looks like that some volcanic eruptions produce cooling and some not. It seems to me that Agung and Pinatubo produced cooling and El Chichon not because of their location. Both Agung (Indonesia, 8S 115E) and Pinatubo (Philippines, 15N 120E) are located in the Indo-Pacific warm pool (the body of water which holds the warmest seawaters in the world):

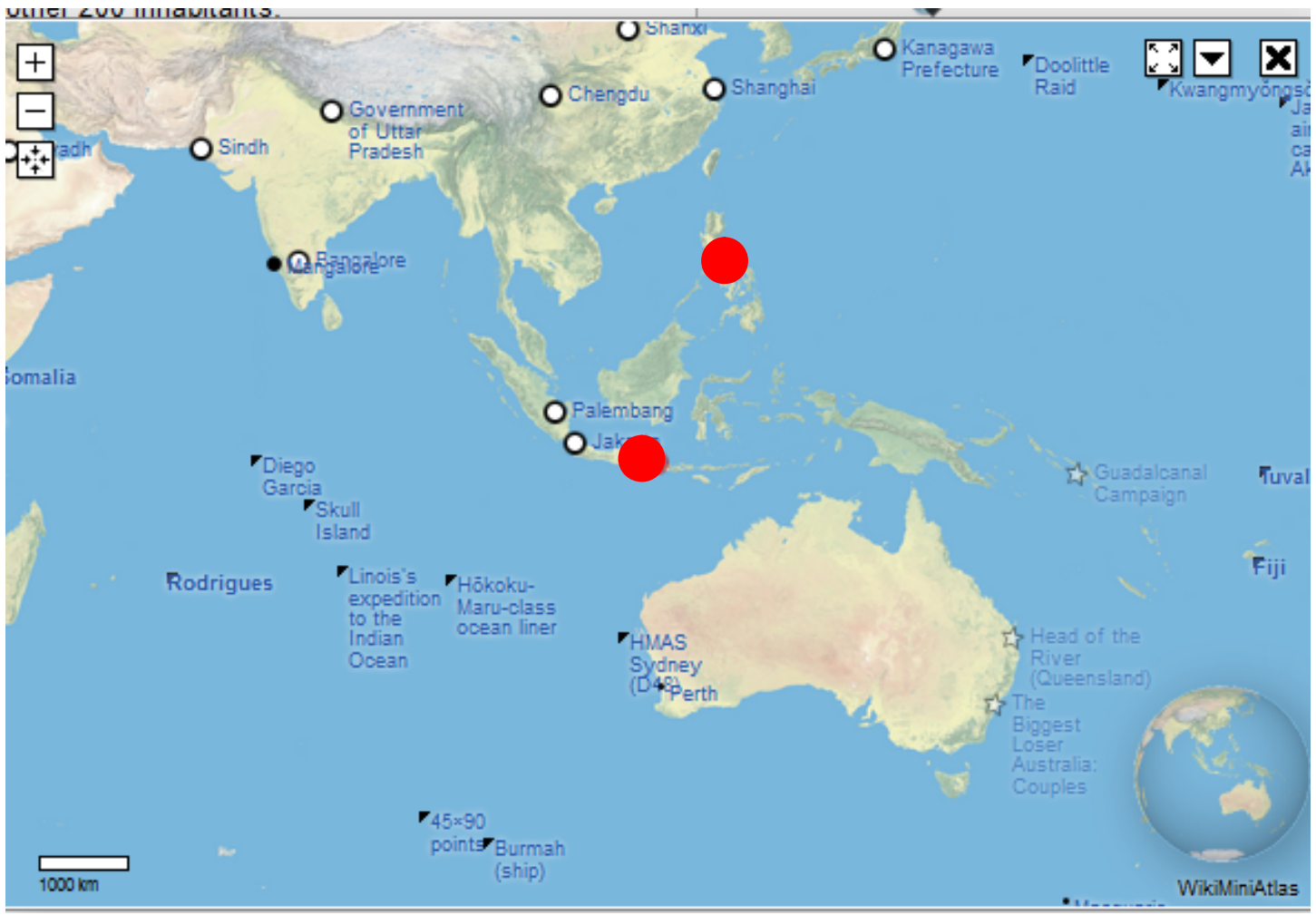

Fig. 15. Agung and Pinatubo locations.

And location of El Chichon(Mexico, 17N 93W) is probably not so good for producing cooling. For example, Mann et. al (2005) wrote "However, the relevant eruptions are the ones that can provide a volcanic dust veil over the tropical Pacific domain of interest... It should be noted that Adams et al. (2003), for this same reason, considered only tropical eruptions in examining the empirical evidence for the response of ENSO to past volcanic forcing."

At last let's look on the plot of tropical (from 20S to 20N) temperature anomalies with linear ENSO effects removed:

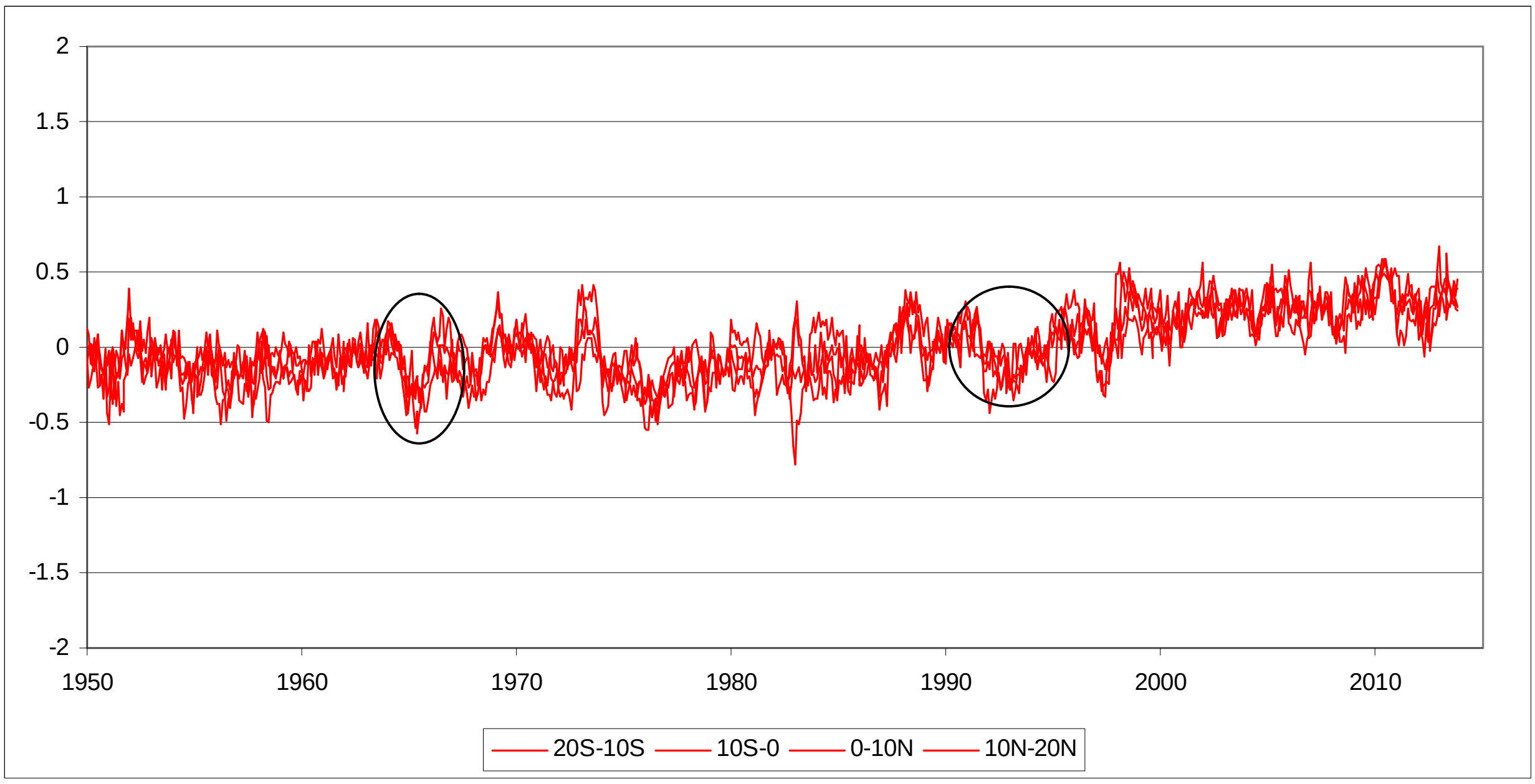

Fig. 16. Temperature anomalies of four ten degrees tropical latitude belts.

Responses to Agung and Pinatubo eruptions are clearly seen and response to El Chichon is near absent. So we have arguments for adjusting to Agung and Pinatubo eruptions. Unfortunately we can't perform the same procedure as for ENSO adjustment. Thus in order to adjust for volcanoes influence we artificially removed downward holes in global time series. And we can't do anything with volcanoes during consideration of spatial patterns.

## Spatial patterns

Then another important question is - if 1987/1988 and 1997/1998 shifts are global or produced by rapid warming in some region. Physics and explanations of the shifts (or the warming) greatly depends on the answer to this question. In order to find the answer let's look on the maps of temperature variations with ENSO effects removed. The maps of linear trends during quasi-stable periods suggest that there are no any global changes and observed variations produced by local changes. According to AGW theory there should be continuous global warming every time modified by local features like AMO at 1950-1987, Pacific heat absorption at 1998-2013. There is no anything like this in trend maps:

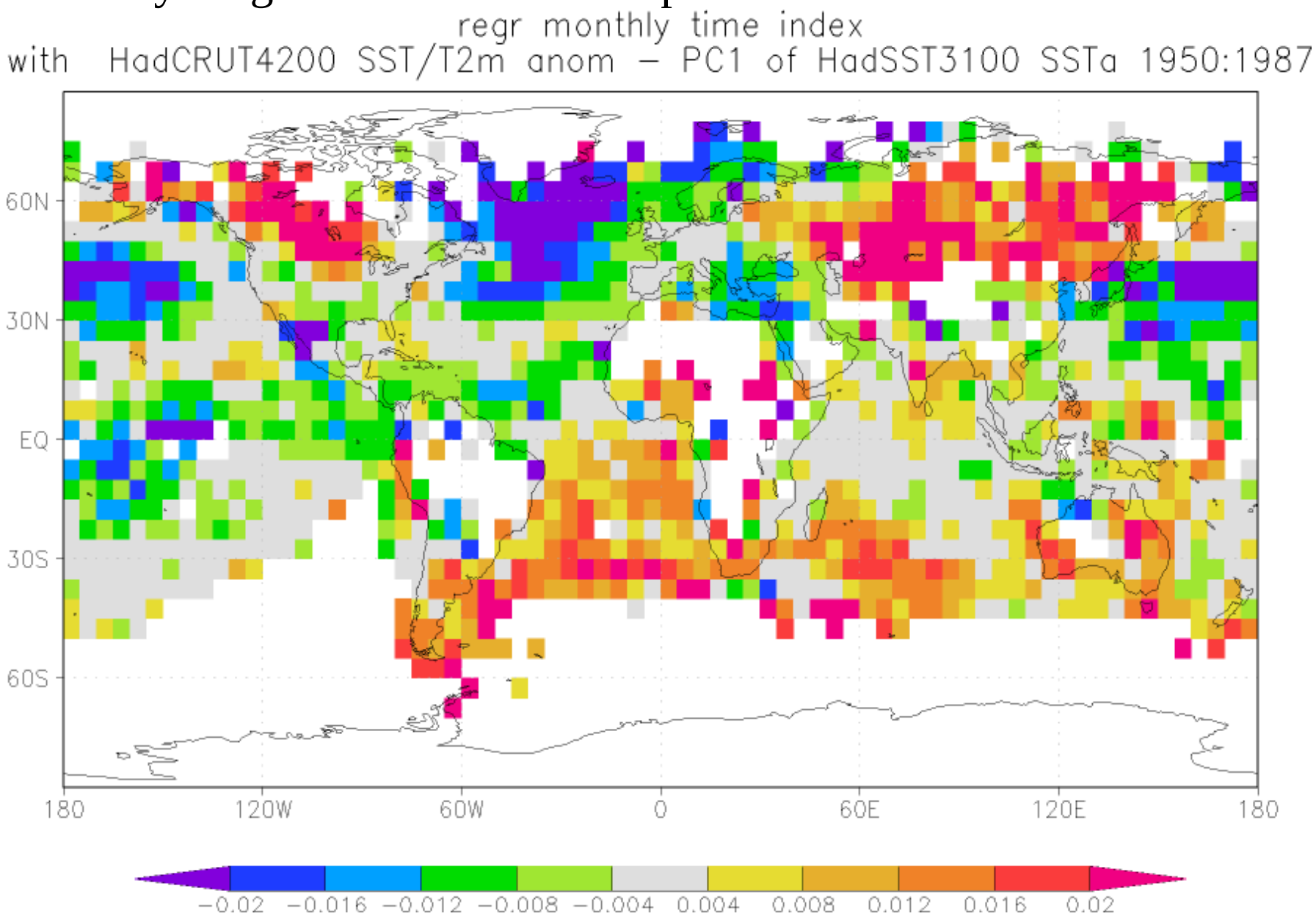

Fig. 17. Map of linear trends of ENSO adjusted HadCRUT4 for 1950-1987 years.

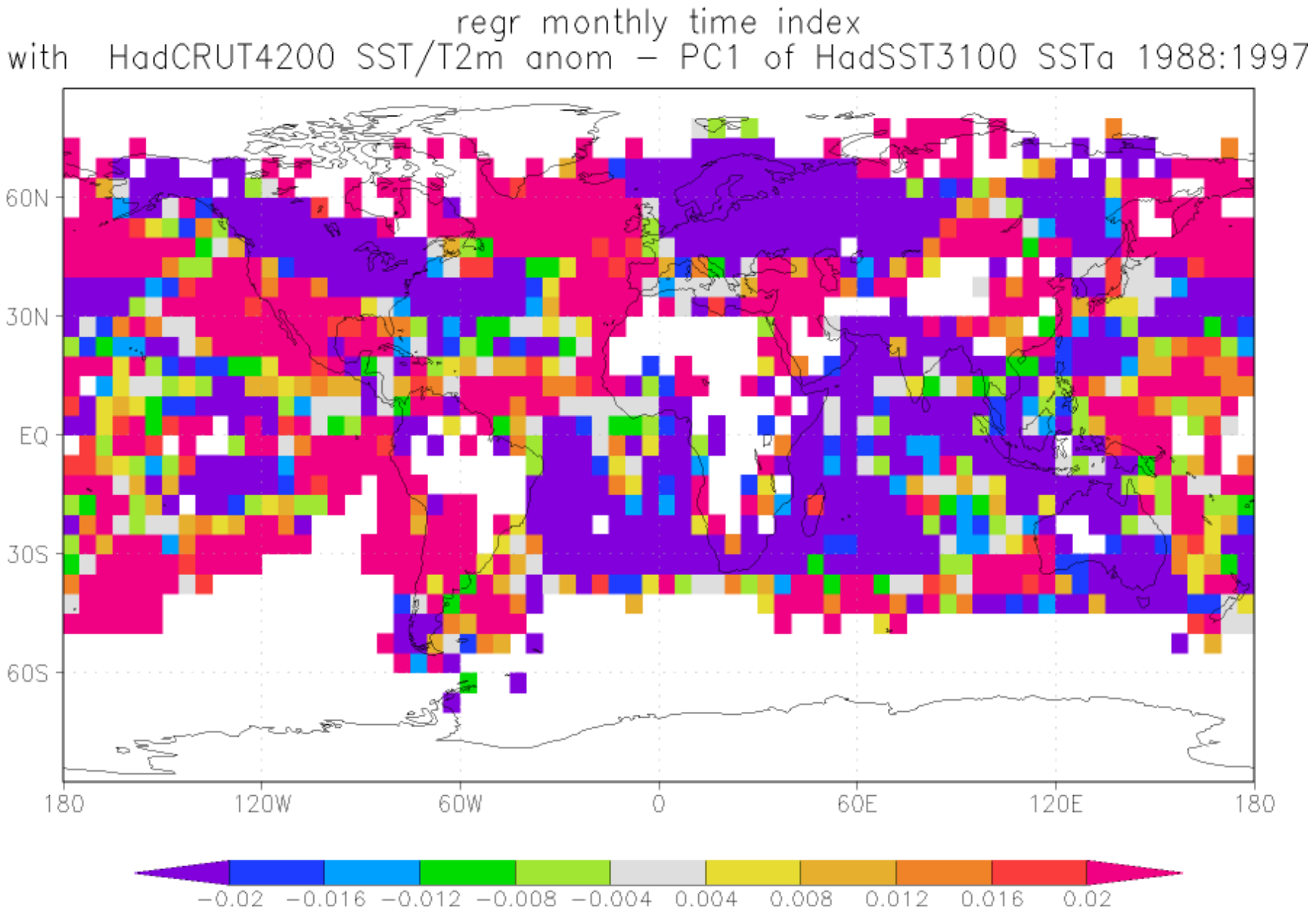

Fig. 18. Map of linear trends of ENSO adjusted HadCRUT4 for 1988-1997 years.

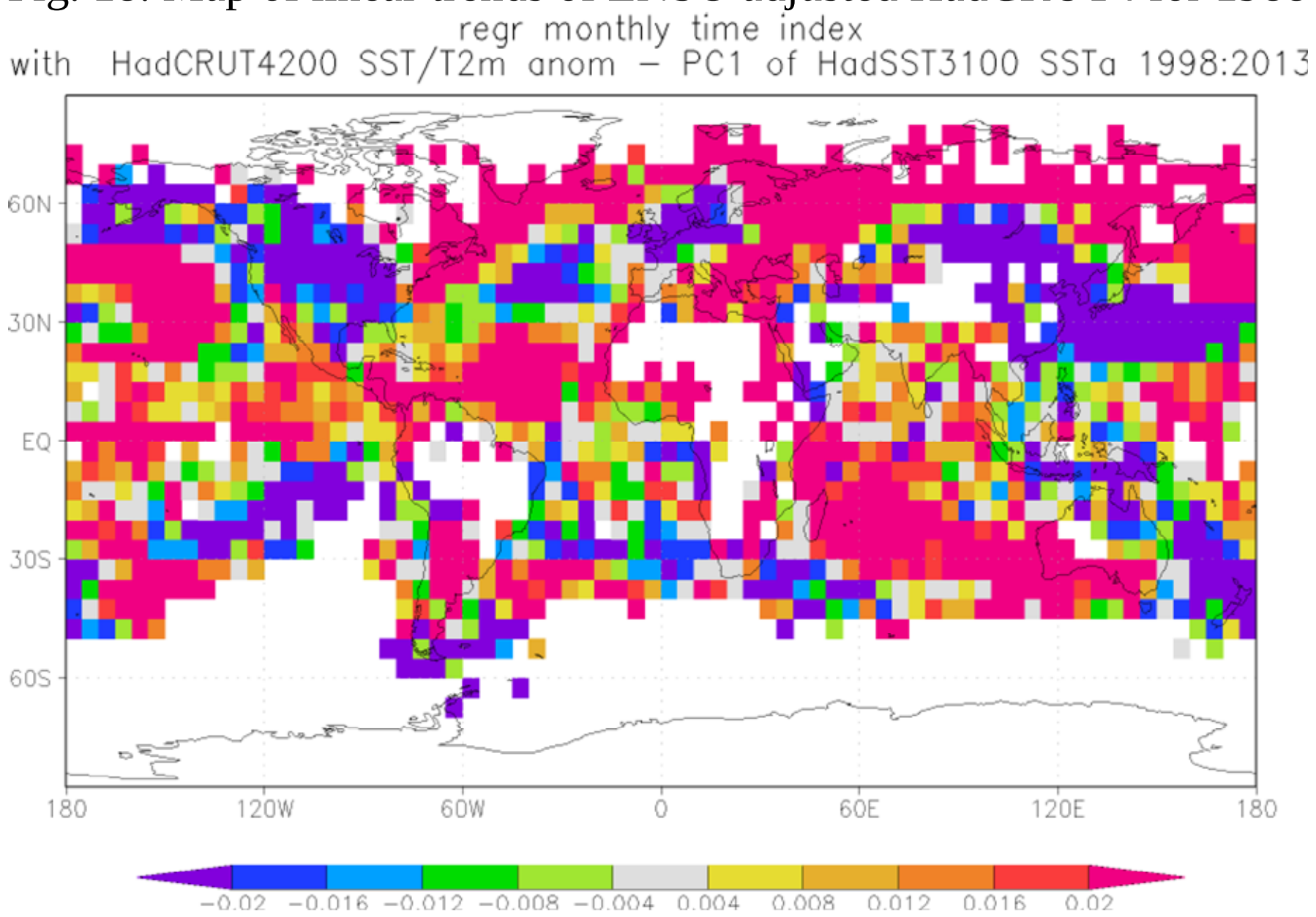

Fig. 19. Map of linear trends of ENSO adjusted HadCRUT4 for 1998-2013 years.

Another question is what places produced rapid growth of global temperature at 1987-1988 and 1997-1998? Let's look on 1988-1991 mean temperature anomalies relative to 1950-1987 mean values and then at 1998-2013 mean relative to 1995-1997 mean. There are still places that are cooler after the shift then before. But at least MAJORITY of regions quickly warms.

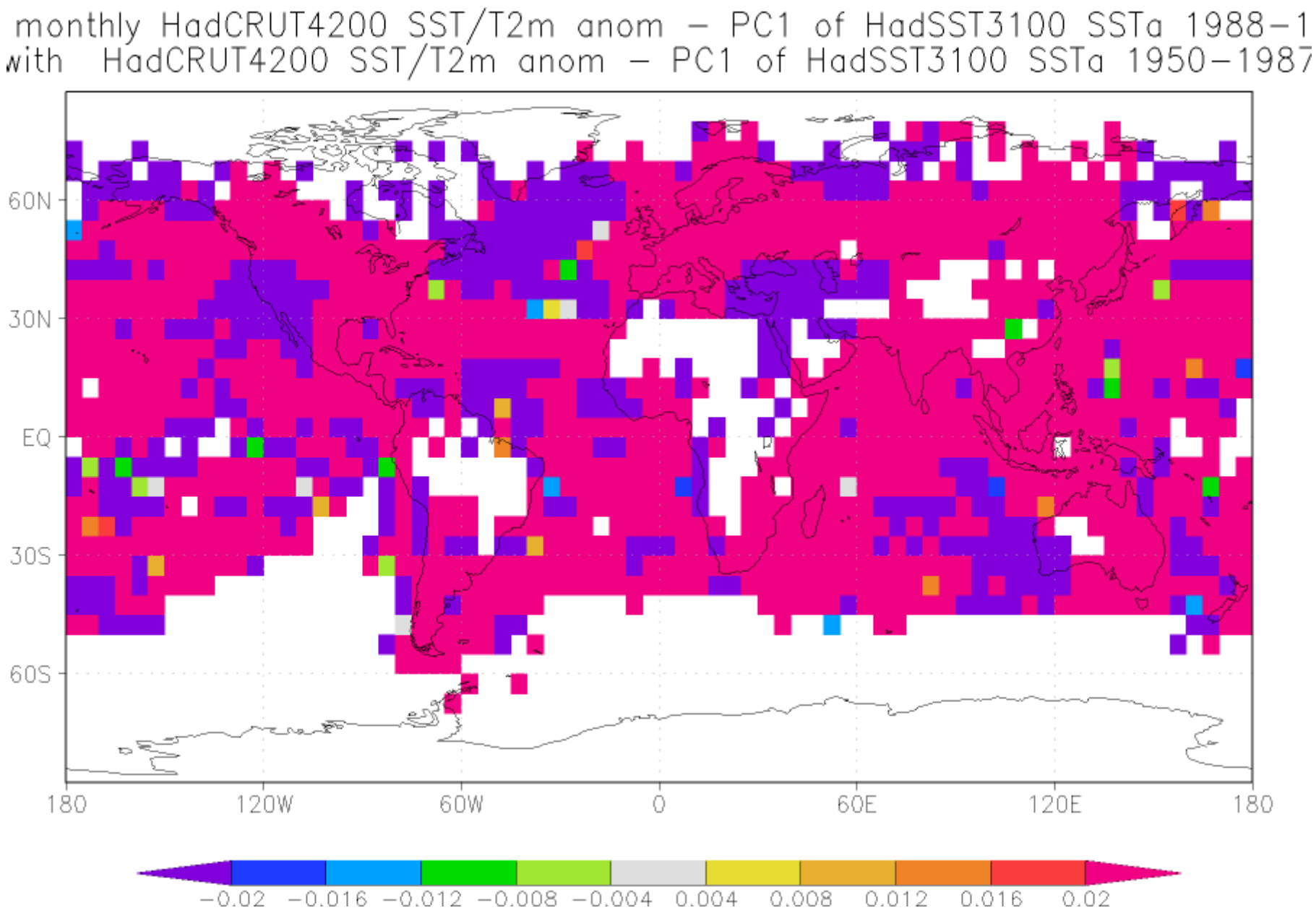

Fig. 20. Map of differences between 1988-1991 and 1950-1987 temperature anomalies of ENSO adjusted HadCRUT4.

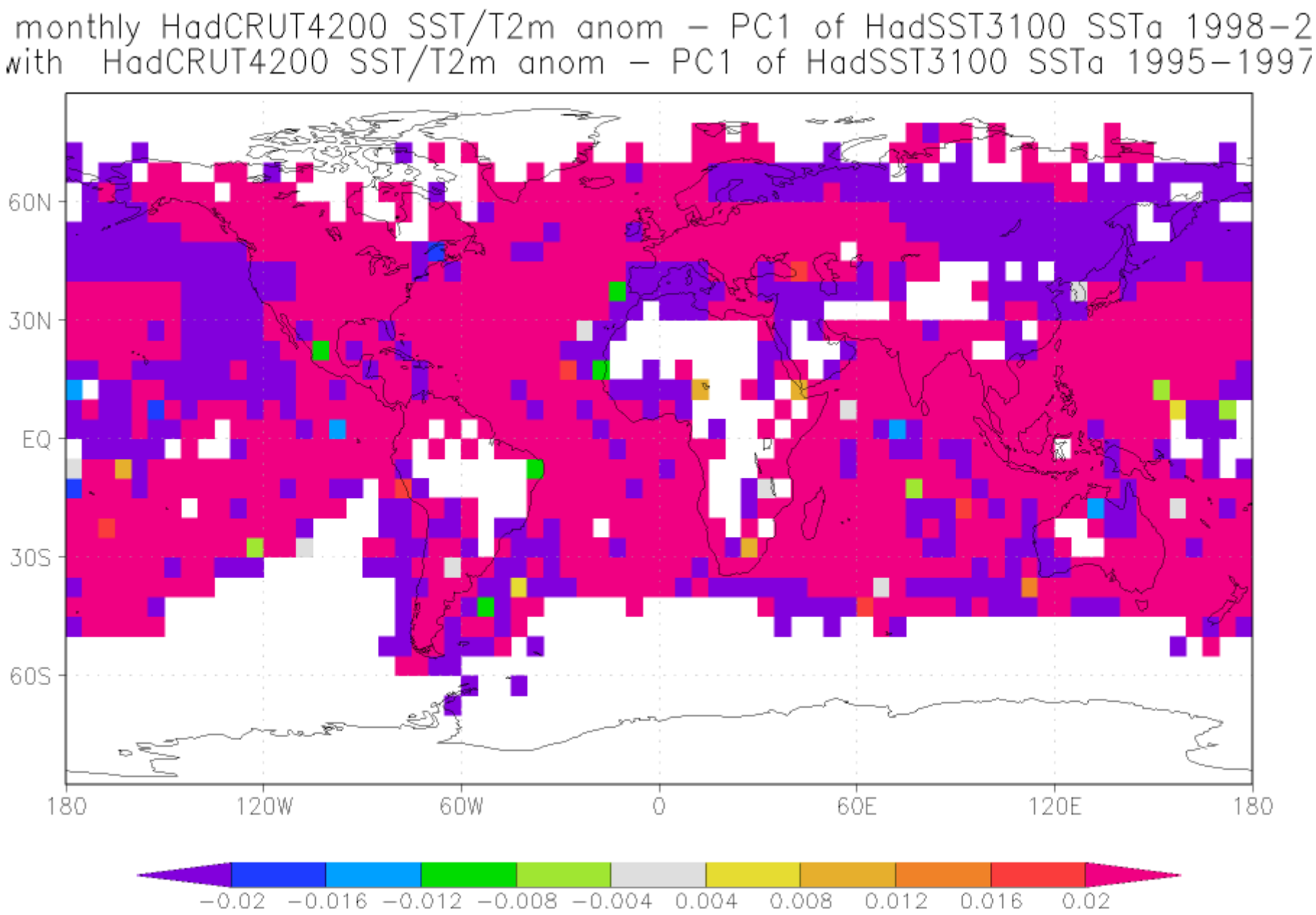

Fig. 21. Map of differences between 1998-2013 and 1995-1997 temperature anomalies of ENSO adjusted HadCRUT4.

## Linear regression model

One of the advantages of our analysis is that it allows to reconstruct adequately global temperatures by only two factors (ENSO and climate regimes). Reconstruction of monthly mean global HadCRUT4 looks so:

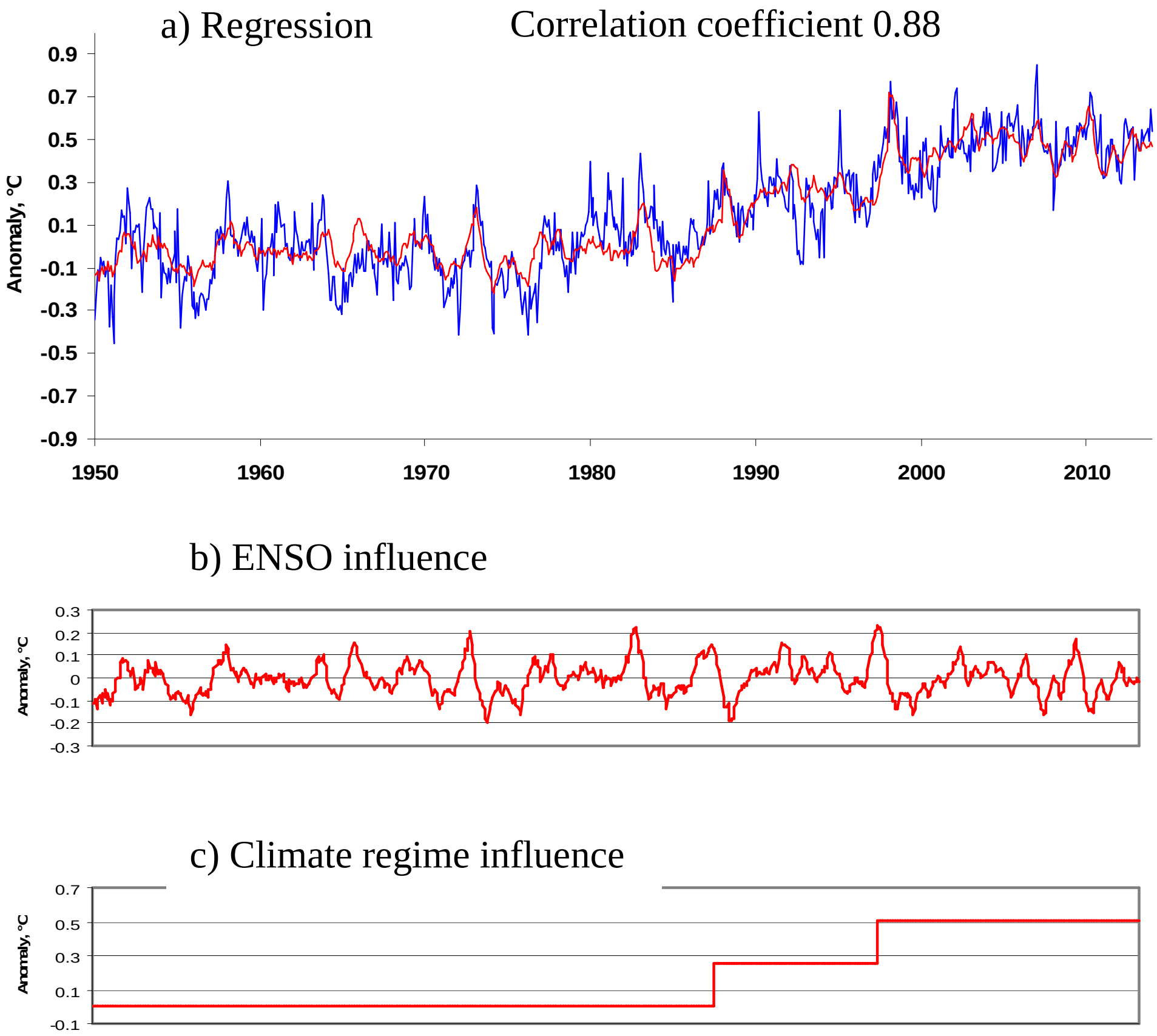

Fig. 22. a) Blue line - HadCRUT4, red line - linear regression on Nino34 and climate regime b) ENSO influence on tropical SST; c) climate regime influence on global temperature anomalies.

## Conclusion

At first, I want to describe how I understand IPCC climate models forcing-response paradigm. They assume that incoming to Earth surface energy balanced by outgoing longwave radiation, which is function of temperature (sigma*T^4). Incoming energy may vary due to changes of Solar input, orbital parameters of Earth and albedo changes. Outgoing energy depends mainly on greenhouse gases concentrations in atmosphere. Also they permit that some intrinsic properties of climate system may influence albedo and outgoing longwave radiation on different timescales - changes of continent positions on geological scale and ENSO variations on the scale of years, for example. So if incoming solar radiation, atmospheric composition and intrinsic properties of climate system are not changing then global mean temperature should be near constant. Anthropogenic greenhouse gases significantly changed atmospheric composition, so according to their logic and calculations (and both logic and calculations at some moment look quite reasonable) this inevitably should lead to global warming. Then there are number of questions - how big should be anthropogenic warming compared to previous climate variations (like Little ice age and Medieval warm period)? What part of observed in 20th century warming is due anthropogenic forcing and what due natural factors? How large observed 20th century warming compared, for example, with Little ice age and Medieval warm period? There are no precise answers and there are many different opinions.

So let's return to IPCC forcing-response paradigm. According to it Earth's global near surface temperature in 20th century passively follows changes in forcing parameters - Solar radiation, atmospheric greenhouse gas concentrations, albedo changes (volcanic and anthropogenic aerosols, black carbon, land use changes). Also there are intrinsic changes or natural cycles of climate system that influence on global temperature - ENSO, PDO, AMO and etc. These natural cycles can influence on incoming or outgoing radiation and modify energy flows between ocean and atmosphere. The picture also become more complicated due to existence of many feedbacks some of which are negative, some positive, and about some it is even hard to say if they are negative or positive. But what is important feedbacks in IPCC climate models may enhance or decrease the response to forcing, but there inevitably should be some response.

Then let's refer to global mean yearly temperature anomalies with removed linear ENSO effects and volcanic eruptions effects from 1950 till now. It should be mentioned that we removed not all ENSO and volcanic eruptions effects. But, what we should prove, we removed most noticeable and really existent effects. So let's look on what we obtained:

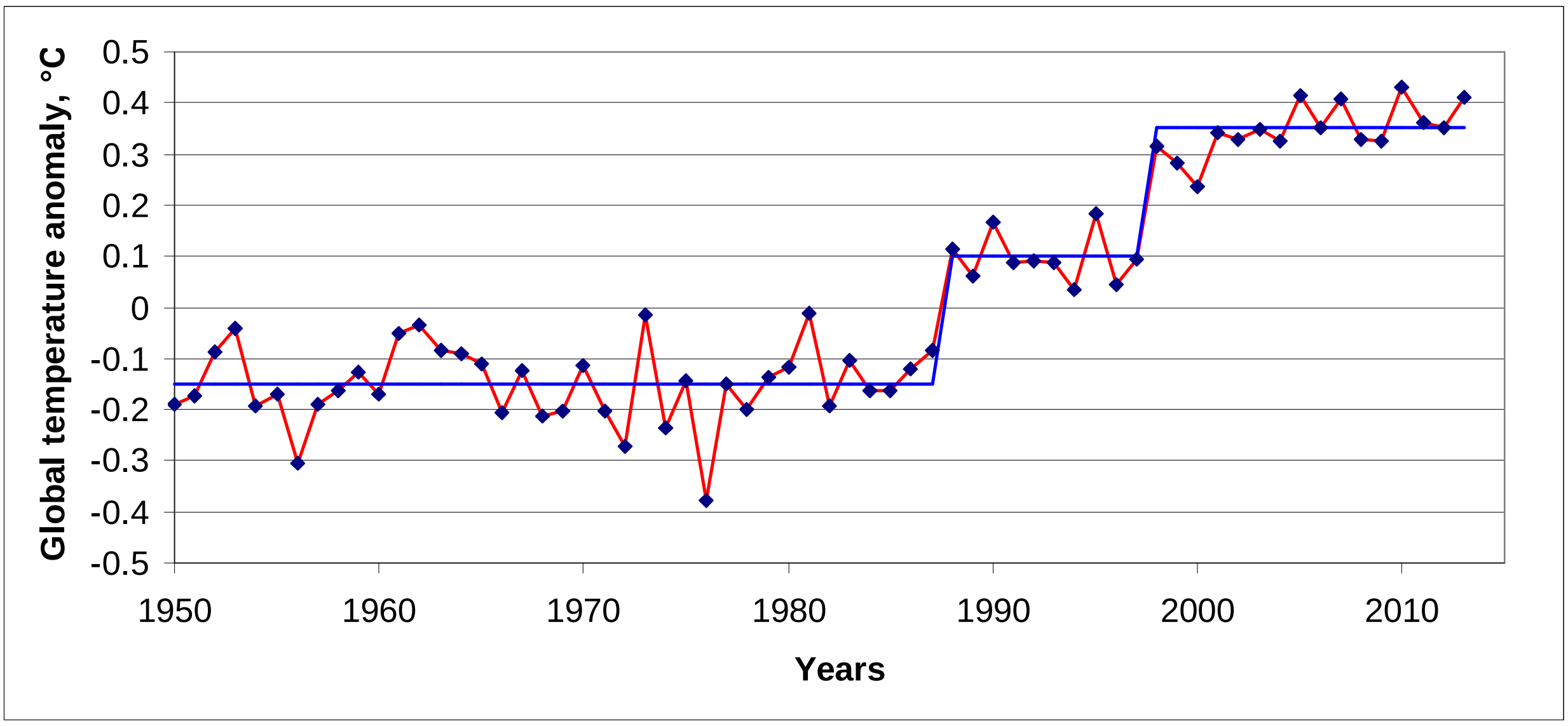

Fig. 23. Yearly averaged global temperature anomalies after ENSO and volcanic adjustment.

Blue line shows our idea about three quasi-stable periods. In this place critics may say that red line (showing observations) is actually not blue line, and there are many points (showing yearly averages) that are quite different from blue line. My reply, is that every time regional properties of climate are changing - for example, Pacific decadal oscillation, Arctic oscillation and so on. Also there are random weather variations that may be important on short timescales. So there is no anything surprising that observations don't lie directly on straight line. But these divergences have no structure, looks random and much smaller by amplitude compared to 1987/1988 and 1997/1998 shifts.

So if we agree that adjusted for ENSO and volcanoes global temperature was near constant for each quasi stable period, 1950-1987, 1988-1997 and 1998-2013, then this phenomena need explanation. And here forcing-response paradigm obviously has problems. According to IPCC there was growing greenhouse forcing during each period. Thus global mean temperatures may stayed near constant only if greenhouse forcing is compensated by other factors. The disagreement between model predictions and observations of global temperature during lasting now "pause" lead to many suggestions of compensating factors - heat absorption by ocean, prevailing La-Nino conditions, PDO transition to cool phase, stadium-waves, aerosols from developing China industry, lowering of solar activity. And, for example, Schmidt et al. (2014) suggest that "a combination of factors, by coincidence, conspired to dampen warming trends in the real world". All this explanations are not impossible. But they are unlikely, because power of compensation factors should grow near synchronically with growing greenhouse forcing. IPCC explanations of the pause already look not so good and they tried to explain only the 1998-2013 stable period. Our analysis add two more periods - 1950-1987 and 1988-1997. Synchrony of changes in greenhouse forcing and compensating factors for each of three periods looks very very unlikely. Then based on the Occam's razor principle, that the simplest explanation is usually the correct explanation, let's search for the simplest explanation of near constant temperature during each period despite growing forcing. **From my point of view the simplest explanation is the existence of some mechanism (thermostat) that regulates global temperature to be near constant despite changing forcing**. I think this is the main result of performed analysis.

I think here we must give precise definitions to what we mean by term "thermostat" and by words "thermostat is operating". In wikipedia thermostat is defined as "a component of a system which senses the temperature of a system so that the system's temperature is maintained near a desired setpoint". From my point of view it is suitable definition. But due to some reasons setpoint of thermostat may change - in this situation we can say that "thermostat is operating". To be more clear, I mean that thermostat can regulate energy flows, so that it can fully compensate additional energy input (from anthropogenic greenhouse gases in our case). But there are another factors that can change the position of setpoint to which temperature is maintained (for example, some of volcanic eruptions) or position of setpoint can change as a result of intrinsic processes - in this case we can say that "thermostat is operating".

Search in the literature showed that the conception of thermostat is not new. The following words were written in the New York Times at 1991: "DOES the earth's climate regulate itself, enabling global temperatures to stay within certain broad limits? The question fascinates scientists, especially at a time of concern and contention about global warming, and now climatologists have produced evidence that clouds act as a natural thermostat that keeps the temperature of the oceans' surface from rising above a certain point. While the evidence comes primarily from the tropical Pacific Ocean, climate experts say it suggests that the earth's closely coupled ocean-atmosphere system could act as a more general thermostat to prevent any warming of the global climate from spiraling out of control." (http://www.nytimes.com/1991/05/07/news/an-oceanic-indication-that-earth-s-climate-might-regulate-itself.html) Thermostat mechanism mentioned here was suggested by Ramanathan and Collins (1991). Later Miller (1997) mentioned many mechanisms which were suggested as thermostats operating in tropics: "Several feedbacks, or ‘‘thermostats,’’ have been proposed to account for the climate stability of tropical convecting regions or the Tropics as a whole, involving, for example, cirrus clouds associated with deep convection (Ramanathan and Collins 1991), upper-tropospheric moisture (Sun and Lindzen 1993b), lateral redistribution of energy by direct circulations (Wallace 1992; Pierrehumbert 1995), surface evaporation (Hartmann and Michelsen 1993), and equatorial ocean dynamics (Dijkstra and Neelin 1995; Clement et al. 1996; Sun and Liu 1996; Seager and Murtugudde 1997). Here we consider whether the tropical climate is stabilized by subtropical low cloud cover." So this idea was in focus of scientific research near twenty years ago but later it was forgotten. And only recently idea of thermostat was advanced by Willis Escenbauch (2009).

Most of suggested thermostats are concerning tropics. But the energy from tropical region distributes to whole globe by ocean currents and atmospheric circulation, so these construction as a whole may act as a global thermostat. This construction may regulate global temperature to be near stable, nevertheless most of regional temperatures are always changing. Then some important parameter have changed during 1987/1988 and 1997/1998 shifts. As a result tropical and whole globe temperatures warmed, but ocean currents and atmospheric circulation continue to vary as usually. I found these explanations based on the long standing global thermostatic hypothesis that has been discussed in scientific literature for several decades more suitable than others.

Here I want to say that many important questions were not considered in this article. Our aim was to present some results that we suppose will be interesting for climate researchers and to invite whose who are interested to take part in developing of these ideas. My e-mail is pbel@icm.krasn.ru.

Finally I want to thank very much Tony Brown for his great concern and support of this study. Also many thanks to Max Anaker, Rud Istvan, Stephen Mosher and Marcia Wyatt for useful comments on yearly version of manuscript.

## Supplementary material

Step by step description of ENSO adjustment

1.) At first you should open Climate Explorer site - climexp.knmi.nl.

2.) Then choose "Monthly *observations*" from "*Select field*" menu:

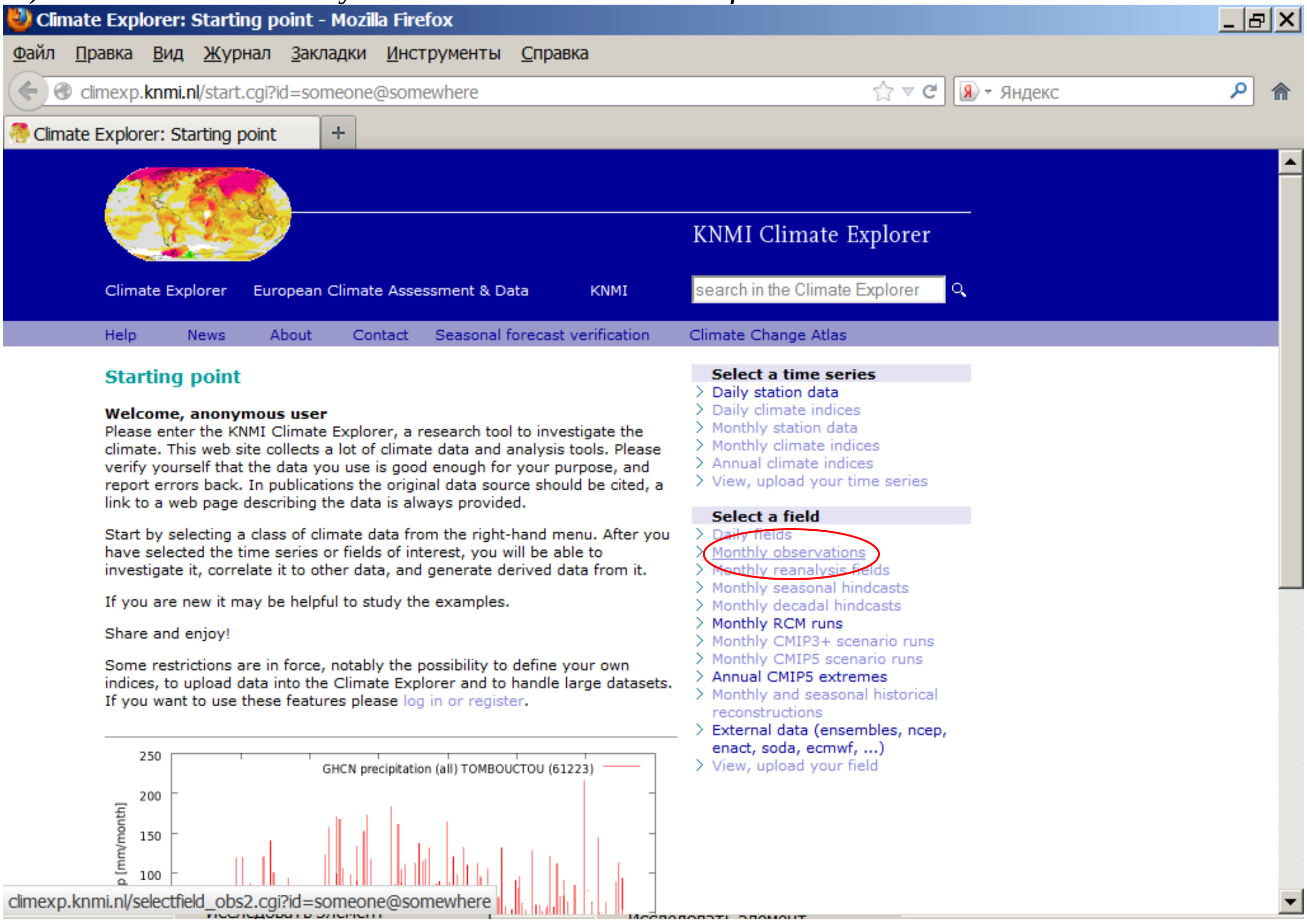

3.) Select HadSST3 field:

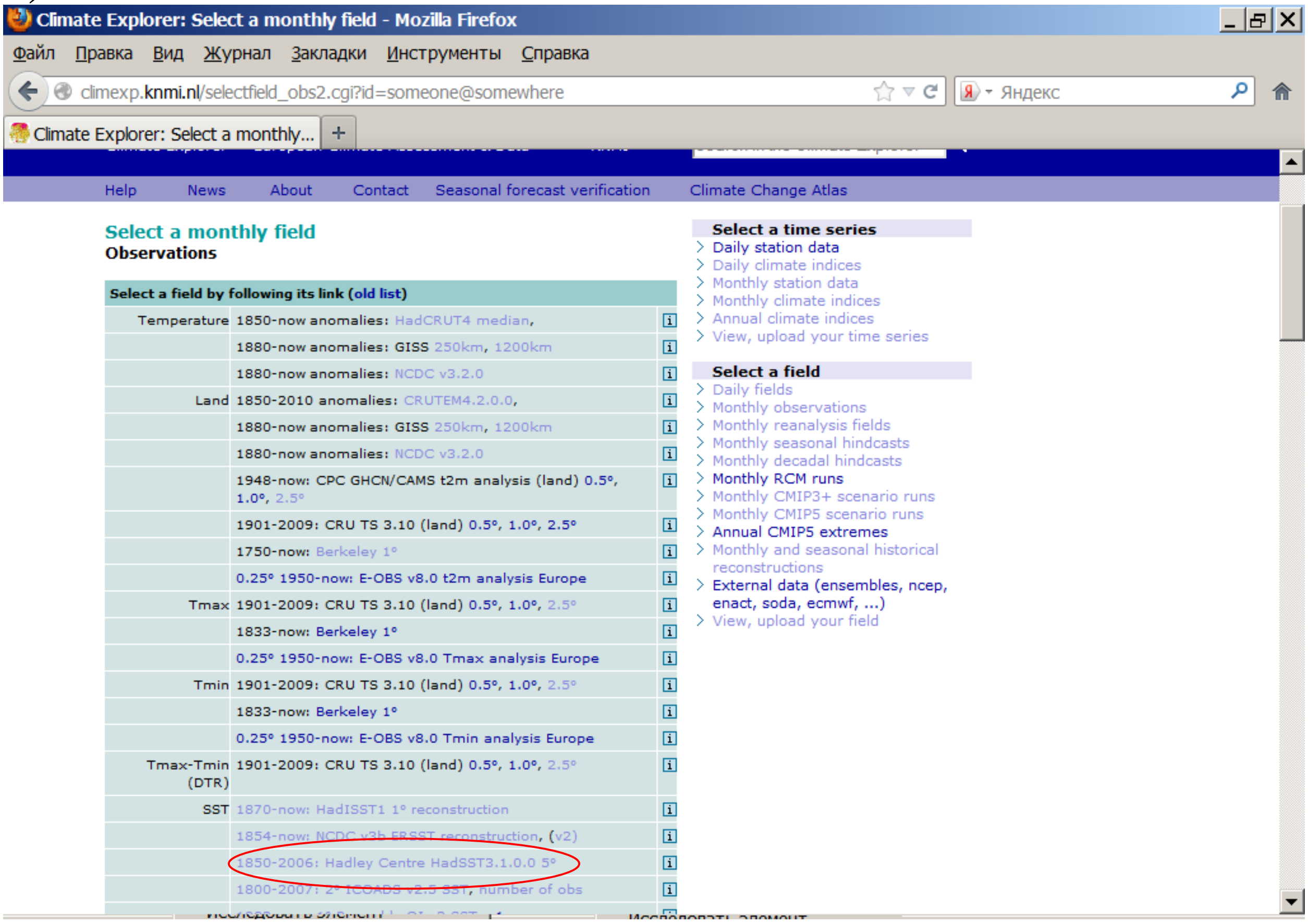

4.) Choose "Make EOFs" from "Investigate this field" menu:

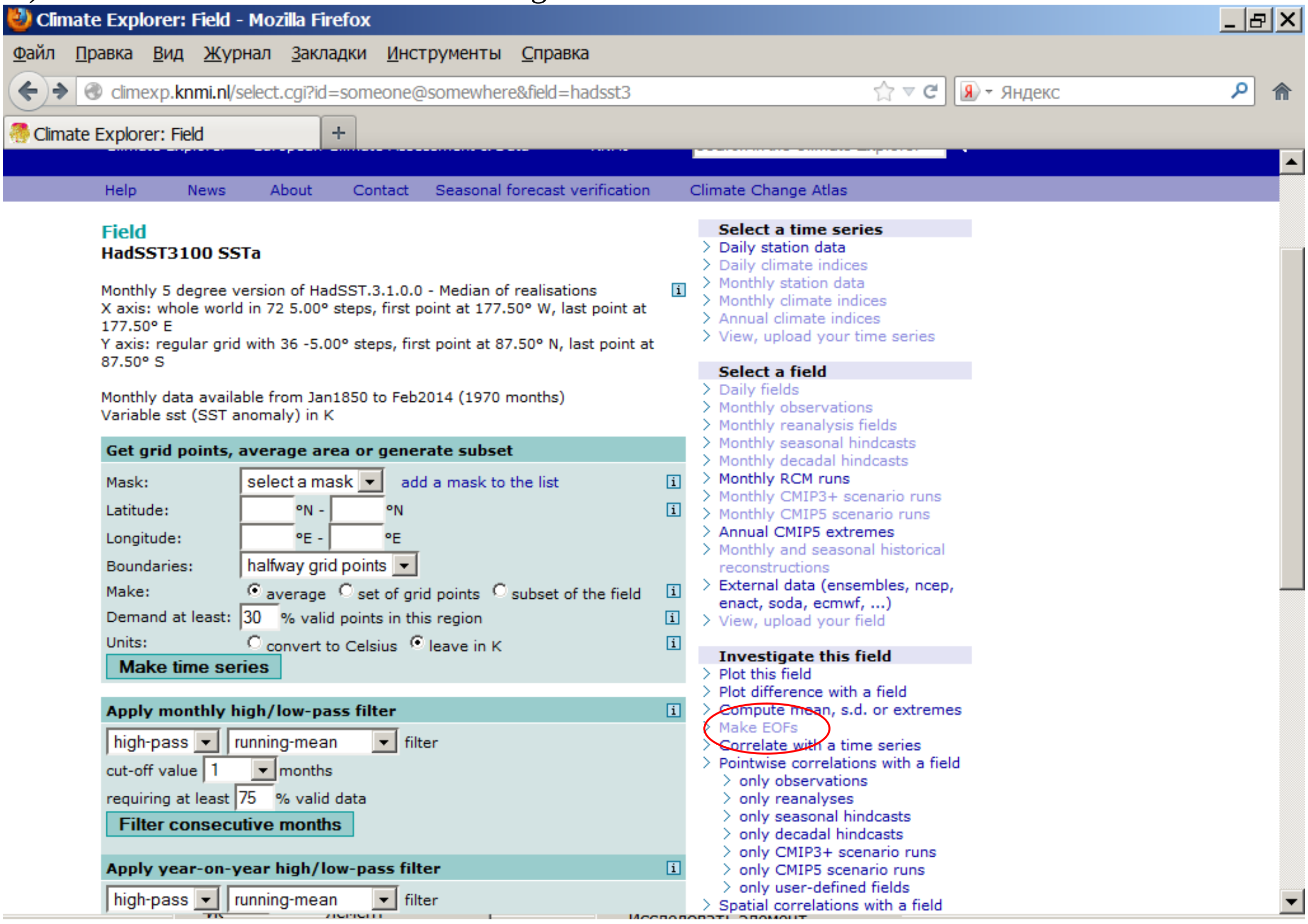

5.) Make several changes to default parameters: choose "together" for "Starting month", 1950-2013 for "Years" and check the box "detrend everything". After that push "Compute EOFs":

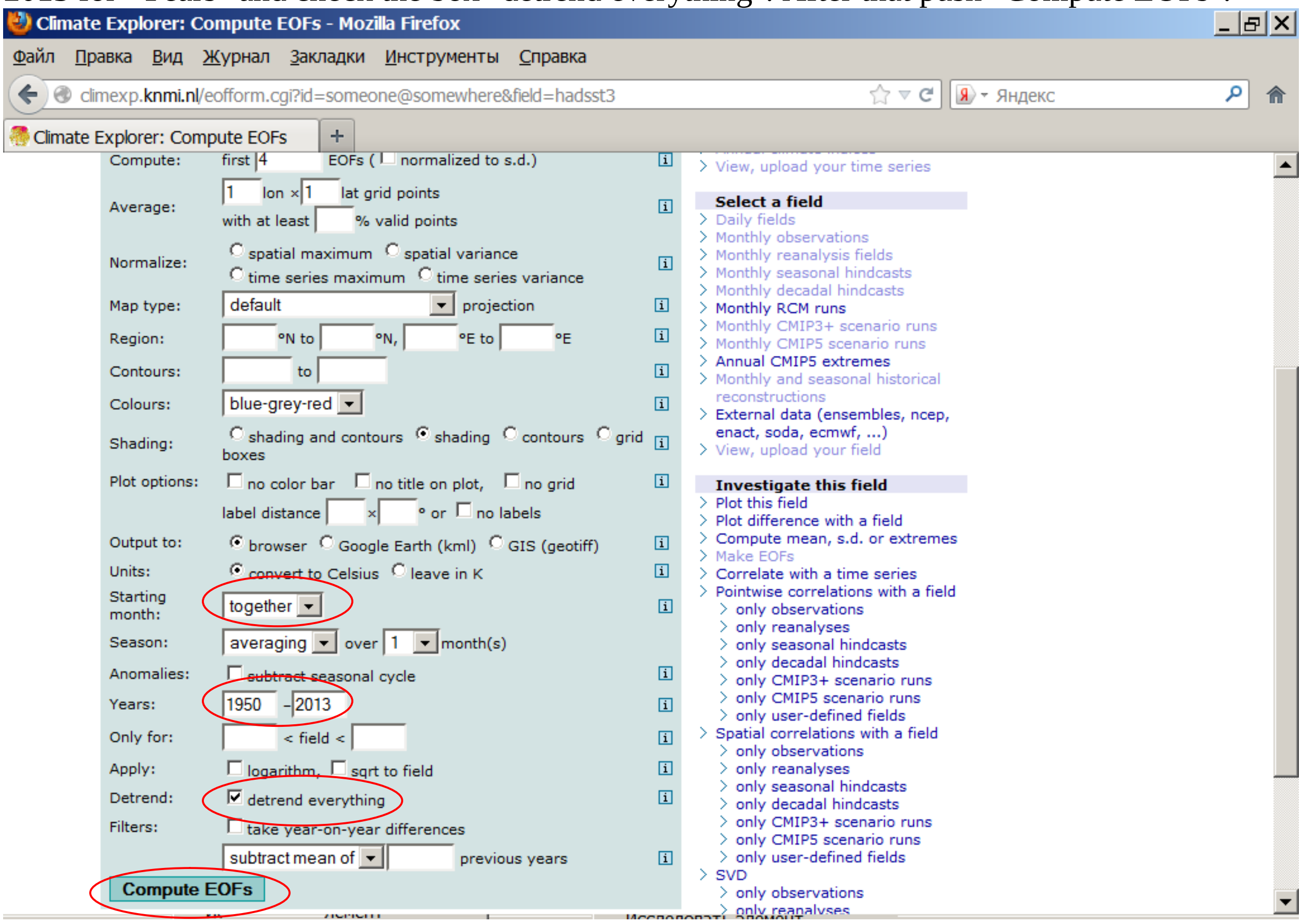

6.) You can see the map of first EOF and how it reproduces the main effects of ENSO. After that choose "Principal component PC1":

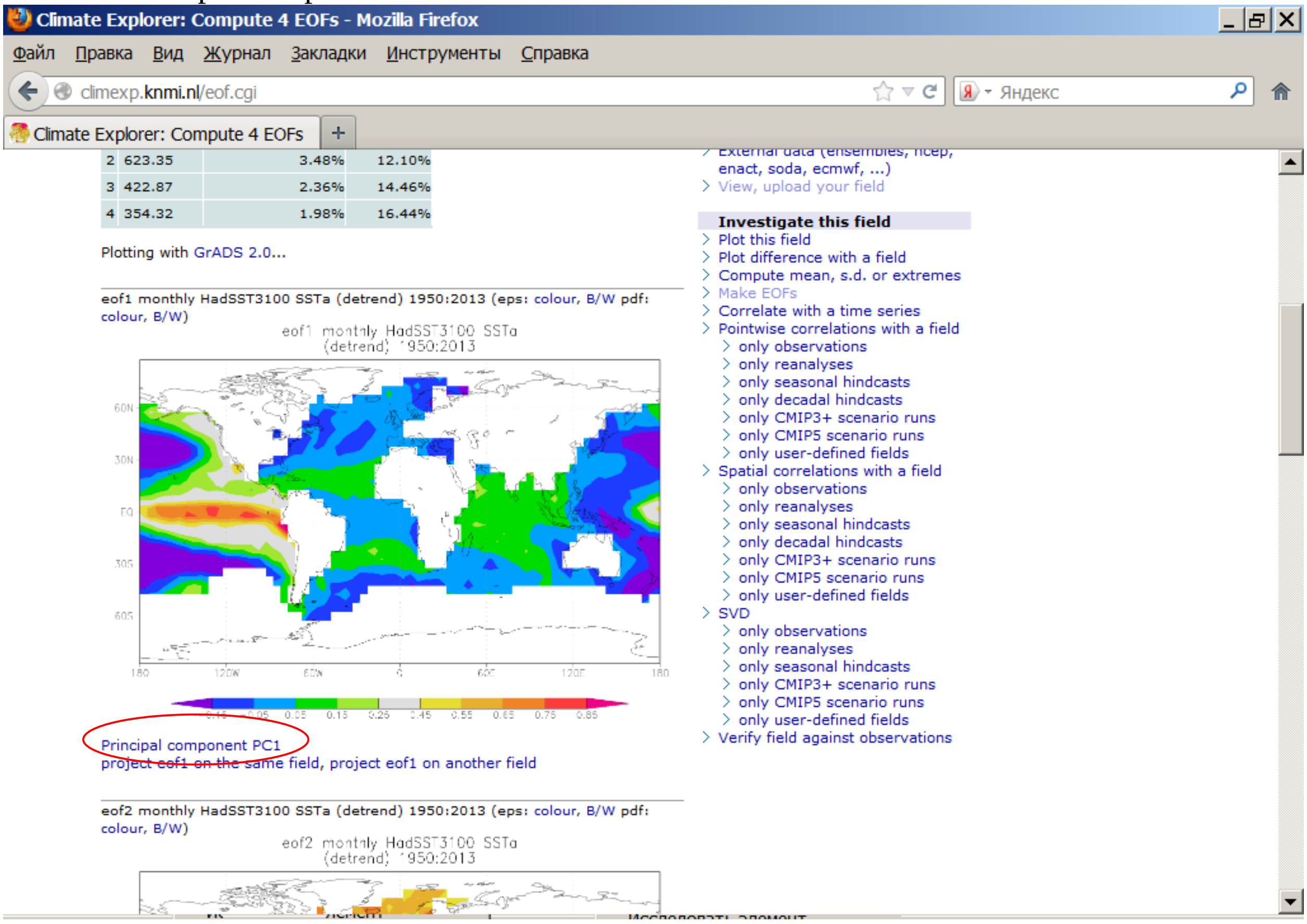

7.) You can see that PC1 is looking near the same as other ENSO indexes, like Nino34 and MEI. It is possible to download calculated time series and perform direct comparison, for example, in Excel. Then choose option "Correlate with a field":

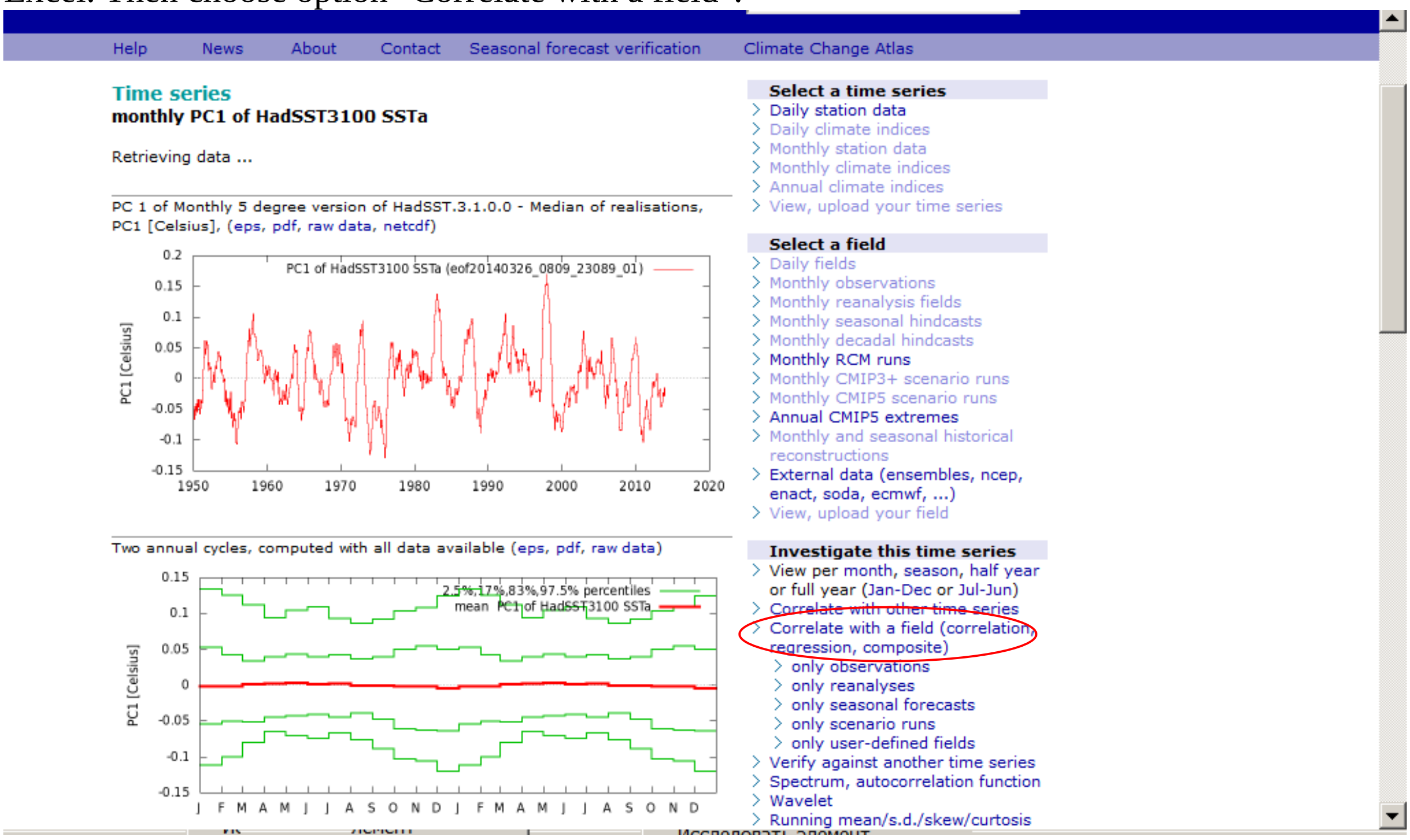

8.) Choose "HadCRUT4 median" and move slider in the bottom to "Options":

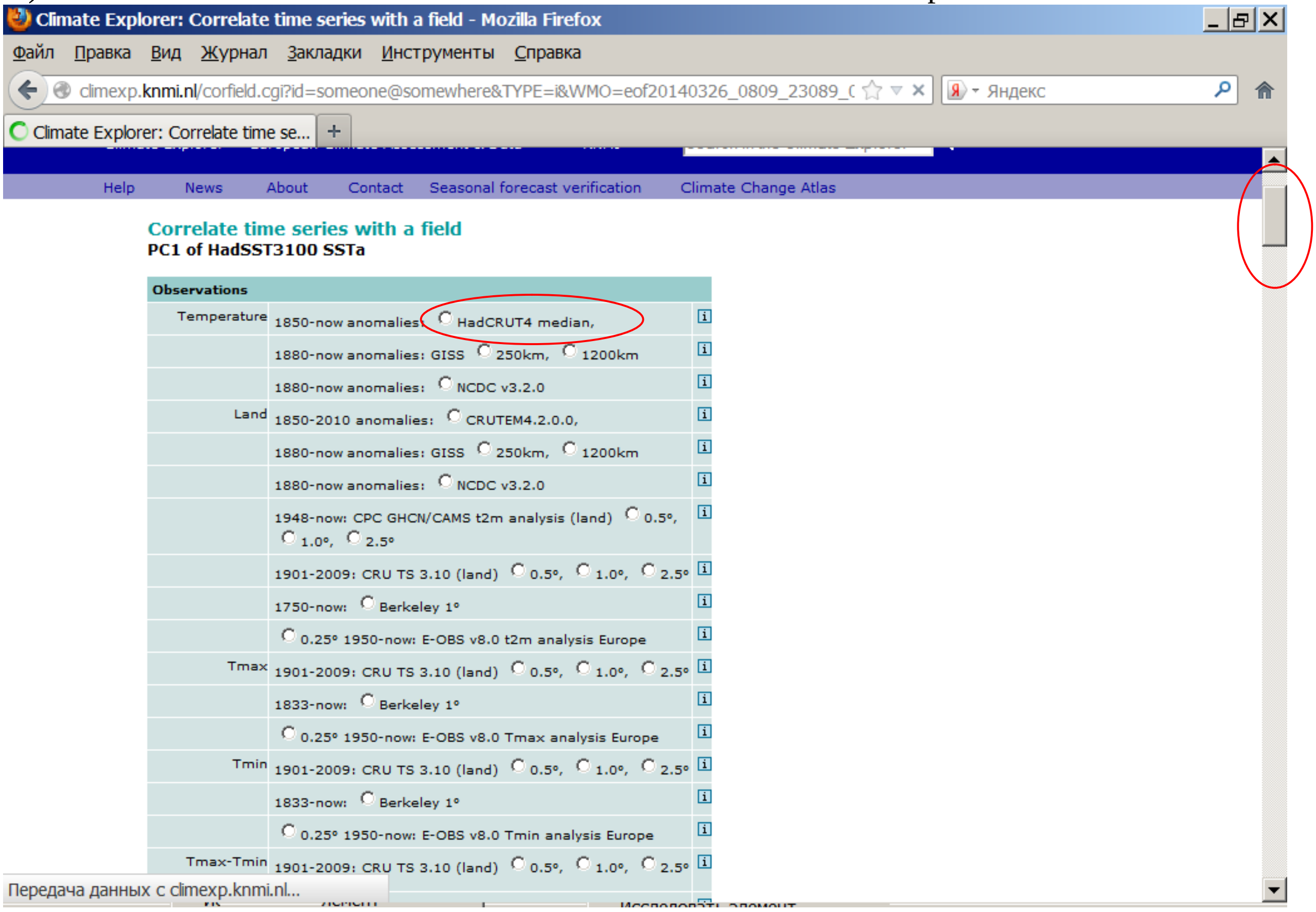

9.) Choose starting month "together", 1950-2013 for "Years", and push "Correlate" button:

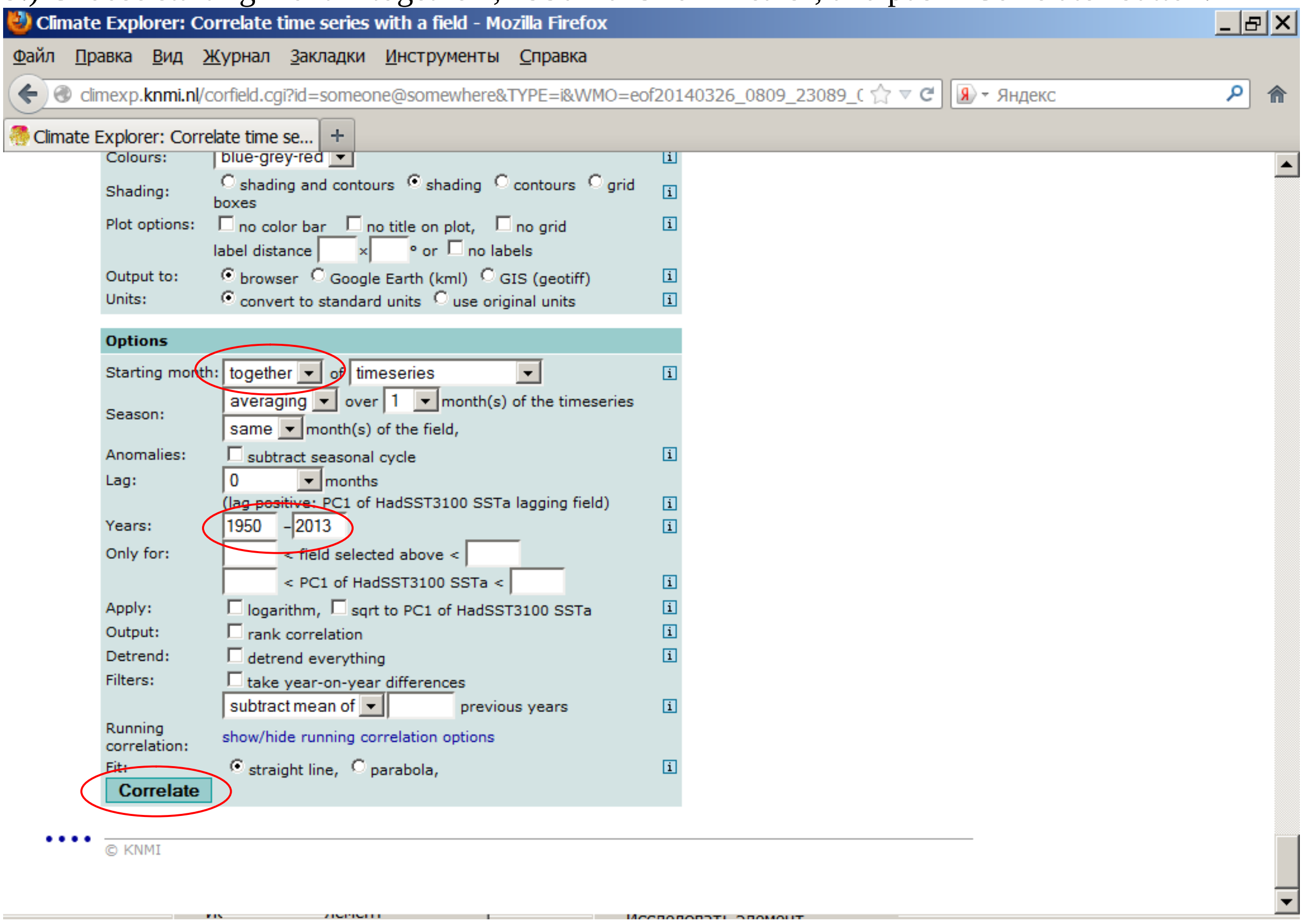

10.) Again you can see effects of ENSO in HadCRUT4 dataset. Then scroll to the bottom of the page:

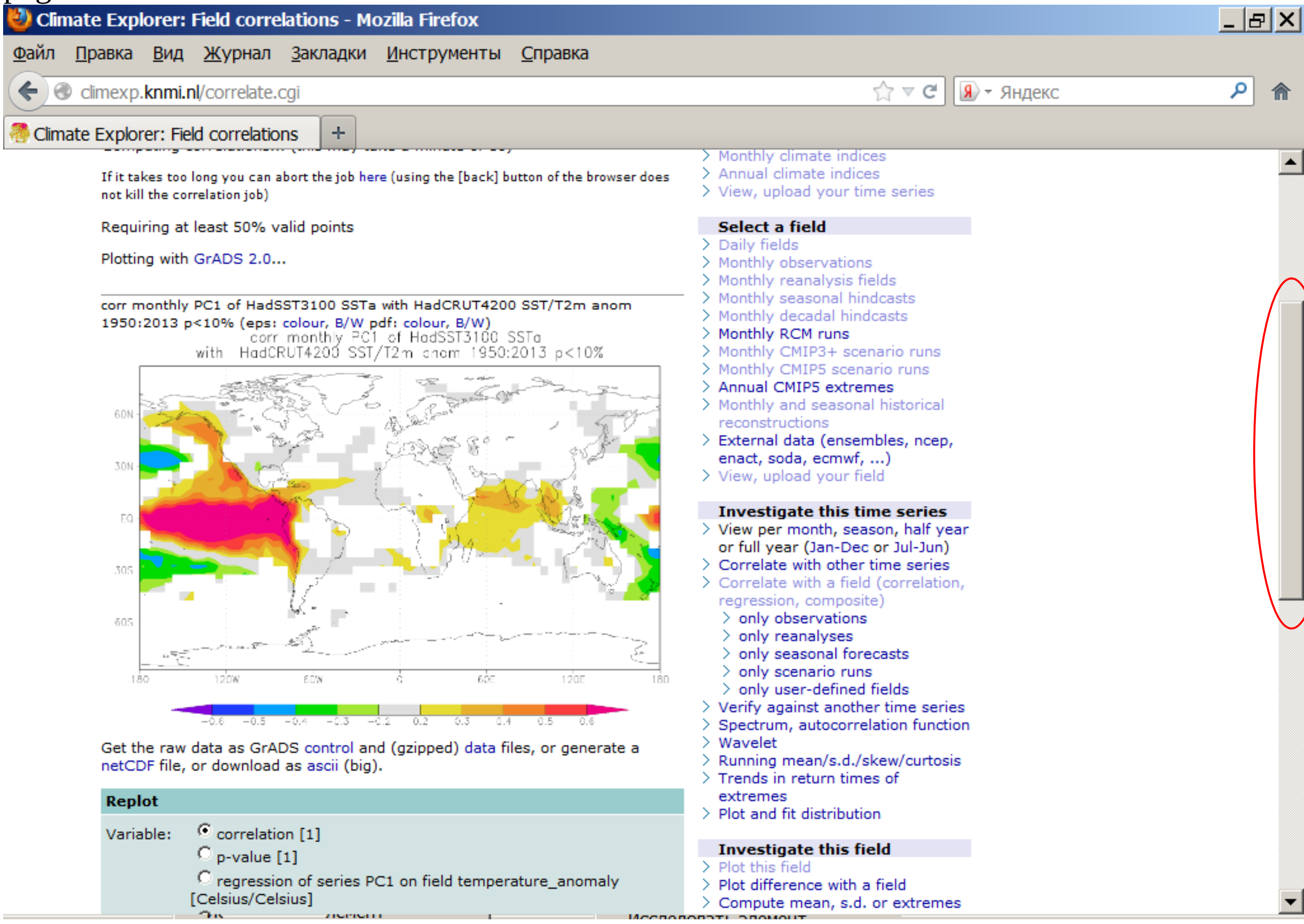

11.) There is the option, that allows to generate a new field with the influence of our ENSO index subtracted linearly. Push "Submit" button:

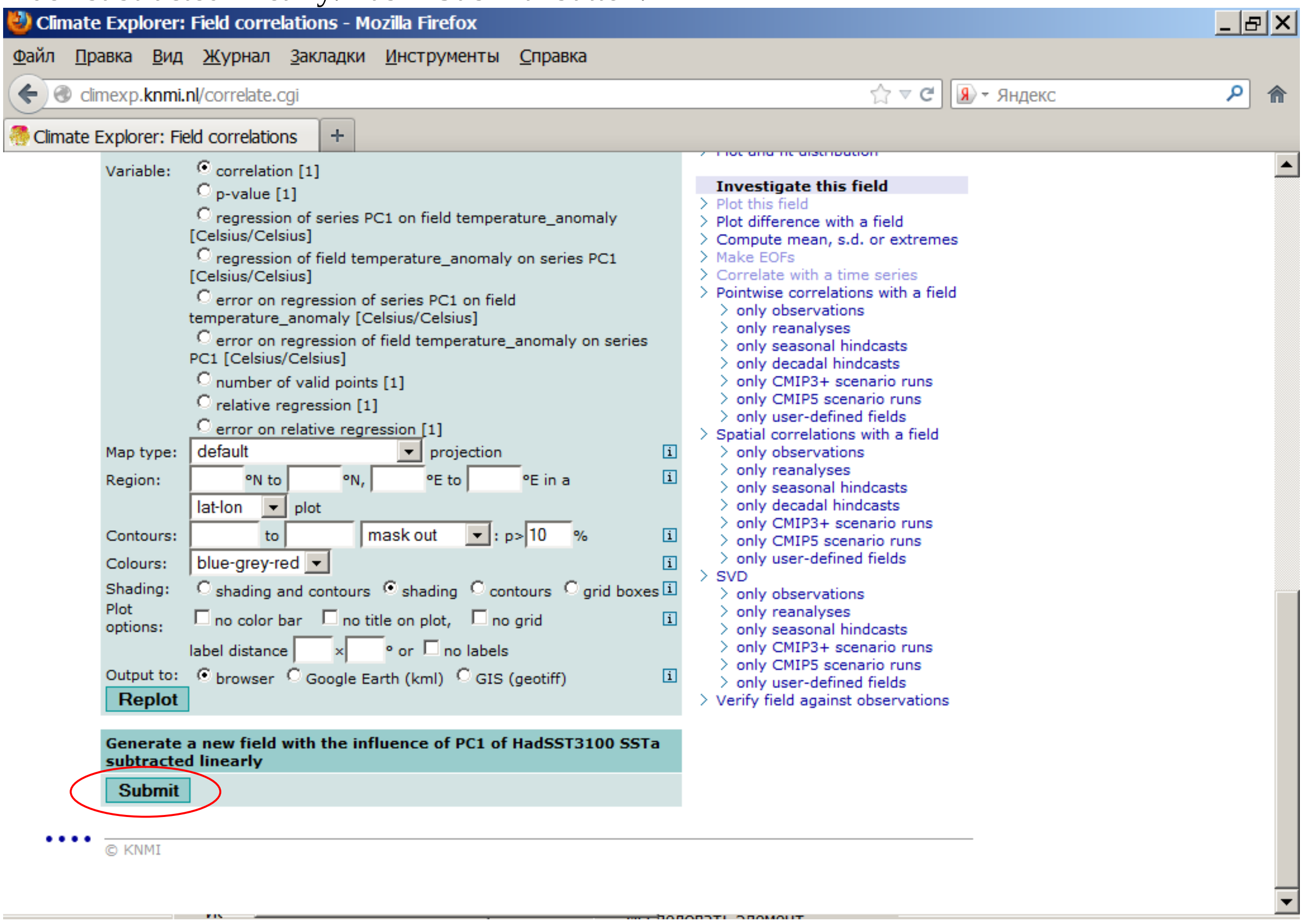

12.) If everything is working good then new field is generated. Now it is possible to use all analysis tools of Climate Explorer as for other fields: plot global and regional temperature anomalies dynamics, plot distribution of trends on a world map and so on. Push button "Continue":

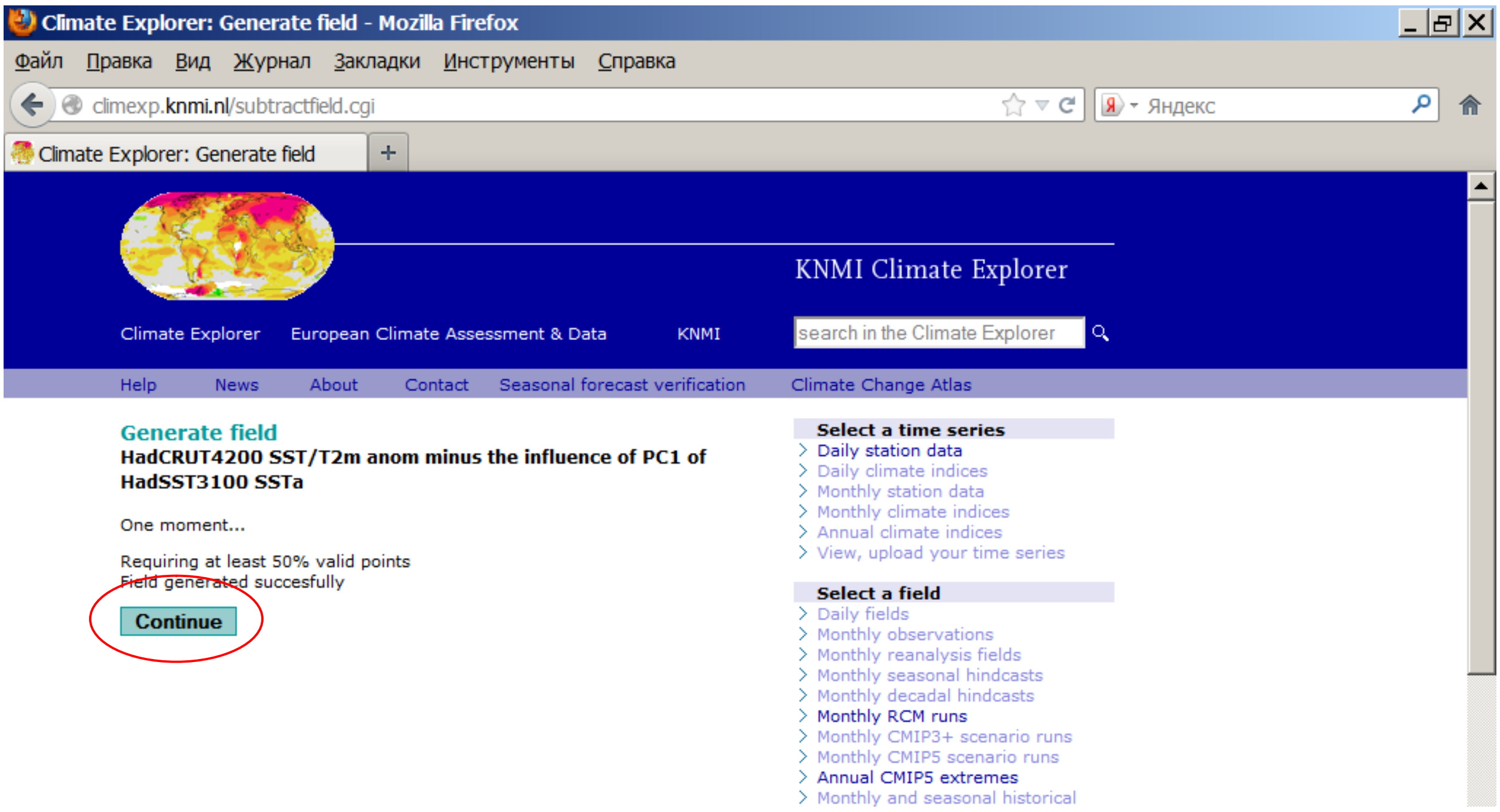

13.) For example, you can now look on ENSO adjusted global mean monthly temperature anomalies. Simply push "Make time series" button:

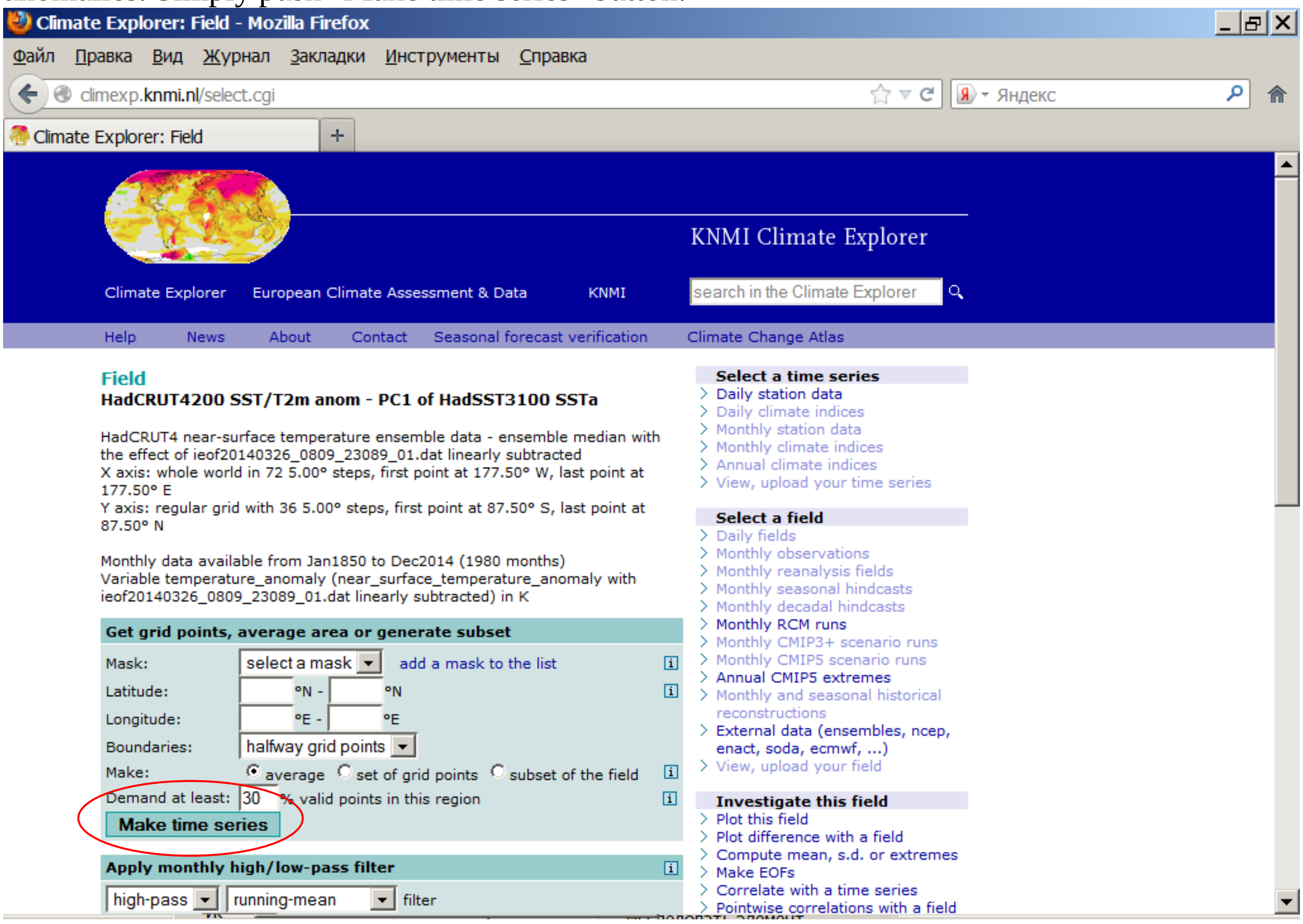

As a result you should obtain the same plot as in the article (with black lines and ellipses of course):

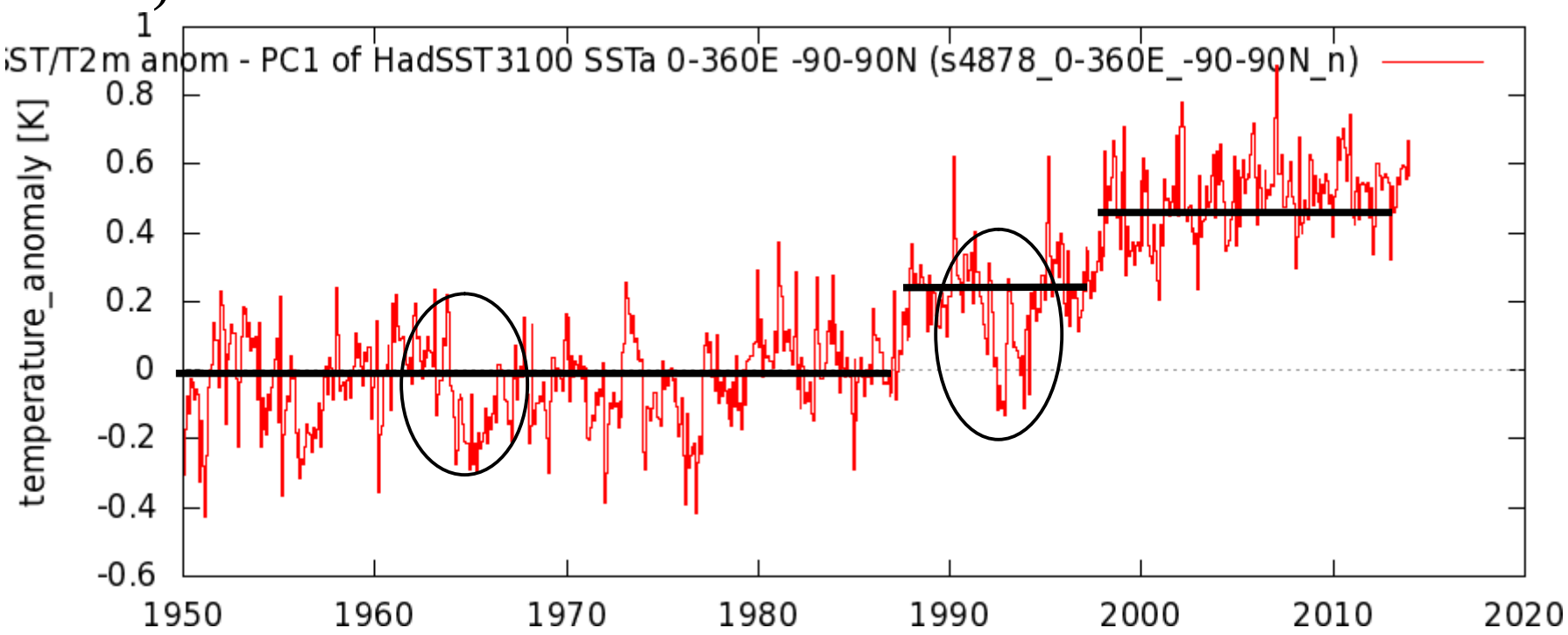